\documentclass[aps,pra,twocolumn,superscriptaddress]{revtex4-2}

\usepackage{graphicx,subfigure}
\usepackage{amsmath,amssymb,bm,amsthm}
\usepackage{mathtools}
\usepackage{multirow}
\usepackage{textcomp}
\usepackage{float}
\usepackage{xcolor}
\usepackage[normalem]{ulem}
\usepackage{dsfont}
\usepackage{url}
\usepackage{mathrsfs}
\usepackage{makecell}
\usepackage{array}

\usepackage{tikz}
\usetikzlibrary{shapes, arrows, positioning}

\newcommand{\ave}[1]{\left \langle #1 \right \rangle}
\newcommand{\ket}[1]{\left | #1 \right \rangle}
\newcommand{\bra}[1]{\left \langle #1 \right |}
\newcommand{\mele}[3]{\left \langle #1 \middle | #2 \middle | #3 \right \rangle}
\newcommand{\inp}[2]{\left \langle #1 \middle | #2 \right \rangle}
\newcommand{\tr}{\text{Tr}}
\newcommand{\opa}{\hat{a}}
\newcommand{\opad}{\hat{a}^\dagger}

\newcommand{\oph}{\hat{H}}

\newcommand{\opp}{\hat{p}}
\newcommand{\opq}{\hat{q}}

\newcommand{\opu}{\hat{U}}
\newcommand{\opud}{\hat{U}^\dagger}

\newcommand{\opn}{\hat{n}}

\newcommand{\oprho}{\hat{\rho}}

\newcommand{\opS}{\hat{S}}

\newtheorem{theorem}{Theorem}
\newtheorem{lemma}[theorem]{Lemma}

\usepackage{hyperref}
\hypersetup{
    colorlinks,
    citecolor=black,
    filecolor=black,
    linkcolor=black,
    urlcolor=black
}

\begin{document}

\title{Passive environment-assisted quantum communication with GKP states}

\author{Zhaoyou Wang}
\email{zhaoyou@uchicago.edu}
\author{Liang Jiang}
\email{liangjiang@uchicago.edu}
\affiliation{Pritzker School of Molecular Engineering, University of Chicago, Chicago, Illinois 60637, USA}

\date{\today}

\begin{abstract}
	Bosonic pure-loss channel, which represents the process of photons decaying into a vacuum environment, has zero quantum capacity when the channel's transmissivity is less than 50\%. Modeled as a beam splitter interaction between the system and its environment, the performance of bosonic pure-loss channel can be enhanced by controlling the environment state. We show that by choosing the ideal Gottesman-Kitaev-Preskill (GKP) states for the system and its environment, perfect transmission of quantum information through a beam splitter is achievable at arbitrarily low transmissivities. Our explicit constructions allow for experimental demonstration of the improved performance of a quantum channel through passive environment assistance, which is potentially useful for quantum transduction where the environment state can be naturally controlled. In practice, it is crucial to consider finite-energy constraints, and high-fidelity quantum communication through a beam splitter remains achievable with GKP states at the few-photon level.
\end{abstract}

\maketitle
% \tableofcontents

\section{Introduction}
One fundamental problem in quantum information theory is to understand the amount of quantum information that can be reliably transmitted through a noisy quantum channel. The bosonic pure-loss channel~\cite{weedbrook2012} describes the ubiquitous phenomenon of energy relaxation in linear systems, where the transmissivity $\eta$ of the channel is the probability that excitations, such as photons or phonons, remain in the system.
For transmissivity $\eta$ more than 50\%, we can protect quantum information against channel noise using bosonic quantum error correcting codes~\cite{grimsmo2020,joshi2021,ofek2016,lescanne2020,grimm2020,michael2016,hu2019,gottesman2001,noh2019,terhal2020,walshe2020,campagne-ibarcq2020,sivak2023,fluhmann2019,deneeve2022,brady2024,albert2018a,wang2022}, such as the cat code~\cite{ofek2016,lescanne2020,grimm2020}, binomial code~\cite{michael2016,hu2019} and GKP code~\cite{gottesman2001,noh2019,terhal2020,walshe2020,campagne-ibarcq2020,sivak2023,fluhmann2019,deneeve2022,brady2024}.
For transmissivity $\eta$ less than 50\%, however, bosonic pure-loss channel is known to have zero quantum capacity~\cite{wolf2007} for direct (i.e., one-way) quantum communication.

\begin{figure}[t]
	\centering
	\includegraphics[width=0.48\textwidth]{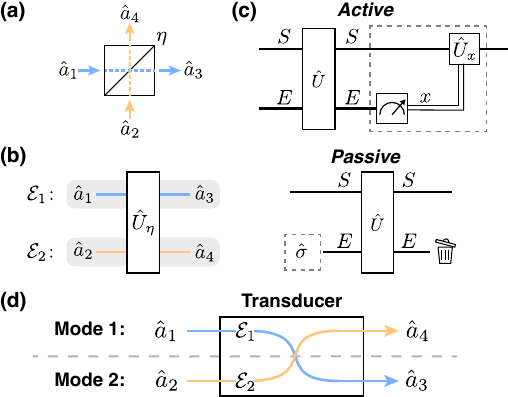}
	\caption{(a) Beam splitter with transmissivity $\eta$. (b) Beam splitter unitary $\opu_\eta$ with quantum channels $\mathcal{E}_1$ and $\mathcal{E}_2$. (c) The performance of a quantum channel acting on an information-carrying system $S$ can be enhanced with active or passive environment assistance. (d) Schematic of a two-mode quantum transducer.}
	\label{fig1}
\end{figure}

The decoherence of a physical system is induced by interactions with its environment, and photon loss in bosonic systems can be modelled as a beam splitter interaction between the system and the environment (Fig.~\ref{fig1}(a-b)). Here $\opa_1(\opa_2)$ and $\opa_3(\opa_4)$ are the input and output modes of the system (environment). The quantum channels $\mathcal{E}_1: \opa_1 \rightarrow \opa_3$ and $\mathcal{E}_2: \opa_2 \rightarrow \opa_4$ describe the evolution of the system and environment.
When the environment mode $\opa_2$ is in a vacuum state, the system evolution $\mathcal{E}_1$ is the bosonic pure-loss channel.

There are several protocols to enhance quantum communication through a quantum channel. A standard approach, known as two-way quantum communication~\cite{pirandola2017}, is to assist the quantum channel with two-way classical communication.
Alternatively, controlling the environment of the quantum channel~\cite{hayden2005,winter2005,memarzadeh2011,lami2020,mele2022,mele2023,oskouei2022} also leads to enhancement, as
every quantum channel can be realized as a unitary interaction $\opu$ between an information-carrying system $S$ and its environment $E$ (Fig.~\ref{fig1}(c)).
Active control involves measuring the environment state after the unitary and applying feedback to the system~\cite{hayden2005,winter2005,memarzadeh2011}.
Both two-way classical communication and active environment assistance enable the transmission of quantum information through bosonic pure-loss channel at arbitrarily small $\eta>0$~\cite{pirandola2017,hayden2005}.
For example, perfect quantum signal transmission through the bosonic pure-loss channel at any $\eta>0$ is achievable with an infinitely squeezed vacuum as the environment input, homodyne measurements on the environment output, and inline squeezing on the system~\cite{zhang2018}.
Passive environment assistance, which we focus on in this work, simply requires choosing some fixed initial state $\hat{\sigma}$ of the environment before the unitary~\cite{lami2020,mele2022,oskouei2022,mele2023} and avoids the need for classical communication.
Table~\ref{table:QC_protocols} classifies existing quantum communication protocols based on their requirements on classical communication and environment control.

For bosonic pure-loss channel, a recent work by  Lami \textit{et al.}~\cite{lami2020} reveals that positive quantum capacities can be achieved even at arbitrarily small $\eta>0$ by utilizing Fock states as the environment state, rather than the vacuum state. This result is surprising, as one might naively expect a vacuum environment to introduce the ``minimal'' amount of quantum noise.
However, the achievable quantum information rates with Fock environment states are only about 0.1~\cite{lami2020}, insufficient for transmitting a single qubit per mode. Moreover, explicit protocols for using a passive environment-assisted quantum channel to transmit quantum information are currently lacking.
Hence, it remains unclear whether better environment states exist beyond Gaussian states~\cite{oskouei2022} and Fock states~\cite{lami2020}, as well as how practically useful the passive environment-assisted protocols can be for realistic applications.

To address these questions, we perform numerical optimization on passive environment-assisted quantum communication, and discover that GKP states emerge as the optimal system encoding and environment state. We prove analytically that perfect transmission of quantum information through the beam splitter is possible for arbitrarily small $\eta>0$ using ideal GKP states.
Additionally, we characterize the entanglement and logical operators of the output states, enabling us to decode the logical information after the quantum channel. This provides a concrete experimental scheme for demonstrating the advantages of environment-assisted protocols.

For practical applications, one crucial requirement is the capability to manipulate the environment state of the quantum channel. Previous works~\cite{lami2020,mele2022,mele2023} have suggested utilizing passive environment assistance in fiber communication, but it requires non-trivial leveraging of the memory effects in optical fibers to control the local environment state. Here, we propose quantum transduction as a potential technology that will benefit from passive environment-assisted schemes. As illustrated in Fig.~\ref{fig1}(d), a quantum transducer effectively performs a beam splitter coupling between the two incoming modes -- one for signal input and the other for passive environment assistance. Since the environments of the transduction channels are naturally accessible, we can improve the performance of quantum transducers with environment assistance. Based on our theoretical findings, direct quantum transduction is achievable at arbitrarily small $\eta>0$ with passive environment assistance.

\subsection{Summary of results}
Our first result is the numerical discovery in Sec.~\ref{sec2:numerical} that near perfect transmission of one logical qubit through a beam splitter is possible at a transmissivity less than 50\%, where the optimized encoding and environment state manifest as GKP states.
In Sec.~\ref{sec2:explanation}, we offer two intuitive interpretations from the perspectives of optical interference and characteristic function to explain the erasure of the logical information from the environment.

\begin{figure}[t]
	\centering
	\includegraphics[width=0.4\textwidth]{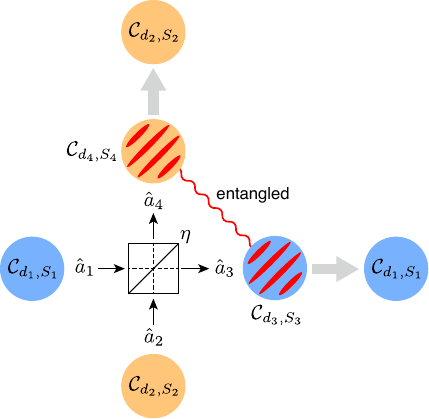}
	\caption{Schematic of the main result. Simultaneous perfect transmission from $\opa_1$ to $\opa_3$ and from $\opa_2$ to $\opa_4$ can be achieved with GKP codes. The gauge subsystems of the two output modes are maximally entangled and shown in red, while the logical subsystems are not entangled.}
	\label{fig_main_result}
\end{figure}

Our main theoretical result is Theorem~\ref{main_theorem} in Sec.~\ref{sec3:main_result}, which constructs the input encodings for achieving perfect transmission through a beam splitter. More specifically, we can construct a $d_1$-dimensional GKP code $\mathcal{C}_{d_1,S_1}$ of mode 1 and a $d_2$-dimensional GKP code $\mathcal{C}_{d_2,S_2}$ of mode 2, and simultaneously transmit them through the beam splitter perfectly (Fig.~\ref{fig_main_result}).

\begin{table}
    \centering
    \begin{tabular}{|l|l|l|l|} \hline
        \textbf{Protocols} & \makecell[l]{\textbf{Classical} \\ \textbf{communication}} & \makecell[l]{\textbf{Environment} \\ \textbf{control}} & $\bm{\eta}$ \\ \hline
        direct QC & No & No & $\eta > 0.5$ \\ \hline
        two-way QC & Yes & No & $\eta > 0$ \\ \hline
        \makecell[l]{active \\ environment- \\ assisted QC} & Yes & Yes & $\eta > 0$ \\ \hline
        \makecell[l]{passive \\ environment- \\ assisted QC} & No & Yes & $\eta > 0$ \\ \hline
    \end{tabular}
    \caption{Quantum communication (QC) protocols categorized by their requirements on classical communication and environment control. The last column represents the required $\eta$ for transmitting quantum information through the bosonic pure-loss channel.}
    \label{table:QC_protocols}
\end{table}

We characterize the entanglement and logical operators for the output states from the beam splitter in Sec.~\ref{sec3:output_states}, and provide several examples  that lead to practical applications including GKP Bell state generation, unidirectional and duplex quantum transduction in Sec.~\ref{sec3:examples}.

We consider finite energy constraints in Sec.~\ref{sec4} and observe that high-fidelity quantum communication through a beam splitter is possible over a wide range of transmissivities using GKP states at the few-photon level.
In Sec.~\ref{sec5}, we introduce the passive environment-assisted quantum transduction which converts quantum signals without classical communication, even in scenarios where the conversion efficiency is less than 50\%.

\section{Optimizing quantum communication through a beam splitter}
\label{sec2}
In this section, we optimize the performance of quantum communication through a beam splitter where GKP states appear as the optimal environment state and encoding for the input quantum signal. We then explain the optimization results from position space wavefunction and characteristic function.

\subsection{Numerical optimization}
\label{sec2:numerical}
The beam splitter unitary $\opu_\eta$ with transmissivity $\eta$ is
\begin{equation}
    \opu_\eta = \exp \left( \arccos \sqrt{\eta} (\opad_1 \opa_2 - \opa_1 \opad_2) \right) ,
\end{equation}
and the input-output relation (Fig.~\ref{fig1}(a-b)) in the Heisenberg picture is
\begin{equation}
	\label{eq:BS_channel}
	\begin{split}
		\opa_3 \equiv & \opud_\eta \opa_1 \opu_\eta = \sqrt{\eta} \opa_1 + \sqrt{1-\eta} \opa_2 \\
		\opa_4 \equiv & \opud_\eta \opa_2 \opu_\eta = \sqrt{\eta} \opa_2 - \sqrt{1-\eta} \opa_1 .
	\end{split}
\end{equation}
Unless otherwise stated, we choose the Schr\"{o}dinger picture for the rest of the paper and use subscripts 3 and 4 to label the output states of $\opa_1$ and $\opa_2$ respectively.
The marginal output states are
\begin{equation}
    \label{eq:marginal_states}
	\begin{split}
		\oprho_3 =& \mathcal{E}_{1} (\oprho_1) = \tr_2 \left[ \opu_\eta (\oprho_1 \otimes \oprho_2) \opud_\eta \right] \\
		\oprho_4 =& \mathcal{E}_{2} (\oprho_2) = \tr_1 \left[ \opu_\eta (\oprho_1 \otimes \oprho_2) \opud_\eta \right] ,
	\end{split}
\end{equation}
with $\oprho_1$ and $\oprho_2$ being the initial states of $\opa_1$ and $\opa_2$.

A passive environment-assisted quantum channel $\mathcal{E}_{\hat{\sigma}}: \mathcal{L}(\mathcal{H}_S) \rightarrow \mathcal{L}(\mathcal{H}_S)$ is defined as (Fig.~\ref{fig1}(c))
\begin{equation}
    \label{eq:env_assisted_channel}
	\mathcal{E}_{\hat{\sigma}} (\oprho) = \tr_E \left[ \opu (\oprho \otimes \hat{\sigma}) \opud \right] ,
\end{equation}
where $\hat{\sigma} \in \mathcal{D}(\mathcal{H}_E)$ is the environment state. Here $\mathcal{L}(\mathcal{H})$ and $\mathcal{D}(\mathcal{H})$ denote the space of linear operators and density operators acting on a Hilbert space $\mathcal{H}$. From Eq.~(\ref{eq:marginal_states}), we know that $\oprho_2$ and $\oprho_1$ are the environment states for $\mathcal{E}_1$ and $\mathcal{E}_2$ respectively.

We can optimize the performance of a quantum channel using entanglement fidelity as the figure of merit, describing how well quantum entanglement is preserved by the quantum channel. 
Entanglement fidelity depends on the encoding and decoding of the input and output states, as well as the environment state for environment-assisted quantum channels.
The dependence is linear and maximization of entanglement fidelity is tractable via iterative convex optimization~\cite{kosut2009,noh2019}.

\begin{figure*}[t]
	\centering
	\includegraphics[width=0.98\textwidth]{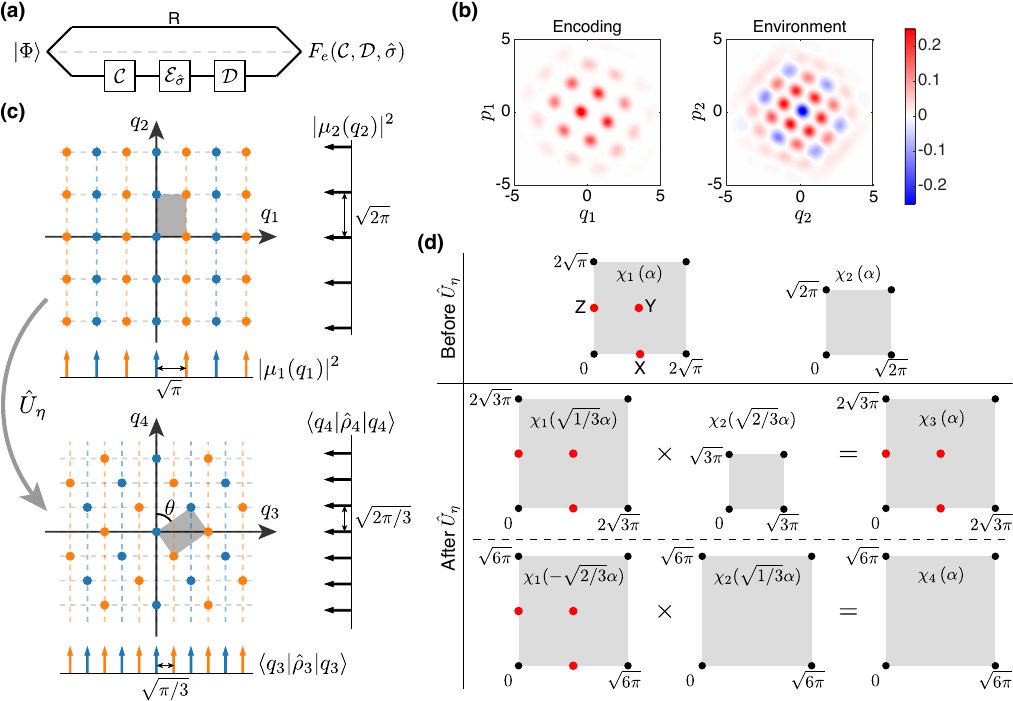}
	\caption{(a) The entanglement fidelity $F_e$ as a function of the encoding map $\mathcal{C}$, the decoding map $\mathcal{D}$ and the environment state $\hat{\sigma}$. (b) Optimized encoding and environment state, which are GKP states with high fidelity. (c) Top: position space wavefunctions of the input states (Eq.~(\ref{eq:square_lattice_GKP_d2}) for mode 1 and Eq.~(\ref{eq:square_lattice_GKP_d1}) for mode 2) before the beam splitter $\opu_\eta$ with $\eta=1/3$. Blue and orange dots represent logical $0$ and $1$. Bottom: output state after the beam splitter where the logical information is erased from $\oprho_4$. (d) The characteristic functions of the input states $\oprho_1$ and $\oprho_2$ before the beam splitter and the marginal output states $\oprho_3$ and $\oprho_4$ after the beam splitter.}
	\label{fig2}
\end{figure*}

Formally, we define the encoding $\mathcal{C}: \mathcal{L}(\mathcal{H}_0) \rightarrow \mathcal{L}(\mathcal{H}_S)$ and decoding $\mathcal{D}: \mathcal{L}(\mathcal{H}_S) \rightarrow \mathcal{L}(\mathcal{H}_0)$, where $\mathcal{H}_0$ is a $d$-dimensional Hilbert space. Given a maximally entangled state on $\mathcal{H}_0$ and a $d$-dimensional reference system $\mathcal{H}_R$ as
\begin{equation}
	\ket{\Phi} = \frac{1}{\sqrt{d}} \sum_{\mu=0}^{d-1} \ket{\mu_R} \ket{\mu} ,
\end{equation}
the entanglement fidelity is defined as (Fig.~\ref{fig2}(a))
\begin{equation}
	F_e (\mathcal{C}, \mathcal{D}, \hat{\sigma}) \equiv \mele{\Phi}{\text{id}_R \otimes \mathcal{D} \circ \mathcal{E}_{\hat{\sigma}} \circ \mathcal{C} (\ket{\Phi} \bra{\Phi}))}{\Phi} .
\end{equation}
Here the $\mathcal{H}_0$ half of the maximally entangled state is sent through the channels $\{ \mathcal{C}, \mathcal{E}_{\hat{\sigma}}, \mathcal{D} \}$, and $\text{id}_R: \mathcal{L}(\mathcal{H}_R) \rightarrow \mathcal{L}(\mathcal{H}_R)$ is the identity map associated with the reference system.
Noticeably, $F_e(\mathcal{C}, \mathcal{D}, \hat{\sigma})$ is tri-linear in $\mathcal{C}, \mathcal{D}$ and $\hat{\sigma}$ and as a result fixing any two of $\{ \mathcal{C}, \mathcal{D}, \hat{\sigma} \}$, optimizing $F_e$ over the remaining variable is a semidefinite programming problem. Therefore we can iteratively optimize the entanglement fidelity over the encoding and decoding maps and the environment state.

We run the optimization for the quantum channel $\mathcal{E}_1$ whose environment state is $\oprho_2$, and find an optimized entanglement fidelity of $F_e \approx 0.98$ for $\eta=1/3$ and $d=2$.
The corresponding coherent information is 0.83, establishing a lower bound on the quantum capacity of the environment-assisted quantum channel.
Therefore passive environment assistance enables us to send one qubit of quantum information with high fidelity through $\mathcal{E}_1$ at $\eta=1/3$. In contrast, the quantum capacity of $\mathcal{E}_1$ is 0 if $\oprho_2$ is a vacuum state since $\eta < 0.5$.

The Wigner functions of the optimized encoding $\oprho_{\mathcal{C}} = \frac{1}{d} \sum_{\mu=0}^{d-1} \mathcal{C} (\ket{\mu} \bra{\mu})$ and the environment state $\oprho_2$ are shown in Fig.~\ref{fig2}(b), which are similar to grid states on a hexagonal lattice.
We fit the optimized states to finite-energy GKP states (Eq.~(\ref{eq:finite_energy_GKP})) with fidelities more than 0.99, and the fitting reveals that the encoding is a $d=2$ GKP code and the environment state is a $d=1$ GKP state. The environment state here is also known as the GKP qunaught state over the hexagonal lattice~\cite{walshe2020}.
Based on the fitting, we set the encoding and environment state as finite-energy GKP states on a hexagonal lattice, and the decoder optimization gives $F_e \approx 0.98$ for $\bar{n} = 3$.

We consistently find GKP states as the optimal encoding across various transmissivities $\eta$, while the environment state is GKP-like for small $\eta$ and approaches a vacuum state as $\eta \rightarrow 1$ up to some random displacements and squeezing (see Appendix~\ref{sec:SI_data}).
Our simulation employs an average photon number constraint $\bar{n} \leq 3$ for the encoding and the environment state with a Fock state cutoff of 20 for both modes. The optimization converges after 150 rounds, where each round comprises 3 iterative steps.

\subsection{GKP states}
Before explaining why GKP states leads to high fidelity quantum communication, we need to first introduce our notations and the GKP states.
For a bosonic mode with creation and annihilation operators $\opa$ and $\opad$ satisfying $[\opa,\opad] = 1$, the quadrature operators are $\hat{q} = (\opa + \opad)/\sqrt{2}$ and $\hat{p} = -i (\opa - \opad)/\sqrt{2}$ where we choose the convention $\hbar=1$. The translation operator in the phase space is defined as
\begin{equation}
	\hat{T}(\bm{u}) \equiv \exp \{i (u_p \opq - u_q \opp) \} ,
\end{equation}
where $\bm{u}=(u_q,u_p) \in \mathbb{R}^2$ is the displacement vector.
Notice that $\hat{T}(\bm{u}) = \hat{D} ((u_q+i u_p)/\sqrt{2})$, where $\hat{D} (\alpha) = \exp (\alpha \opad - \alpha^* \opa)$ is the usual displacement operator. The commutation relation is
\begin{equation}
	\hat{T}(\bm{u}) \hat{T}(\bm{v}) = e^{-i \omega (\bm{u},\bm{v})} \hat{T}(\bm{v}) \hat{T}(\bm{u}) ,
\end{equation}
where the phase
\begin{equation}
	\omega (\bm{u},\bm{v}) = u_q v_p - u_p v_q = \det \begin{pmatrix} u_q & u_p \\ v_q & v_p \end{pmatrix} 
\end{equation}
is the oriented area of the parallelogram spanned by vectors $\bm{u}$ and $\bm{v}$.

GKP states~\cite{gottesman2001} are defined as states that are invariant under certain phase space displacements.
Given two displacement operators $\hat{S}_X = \hat{T}(\bm{u})$ and $\hat{S}_Z = \hat{T}(\bm{v})$, the states that are invariant under the stabilizers $\hat{S}_X$ and $\hat{S}_Z$ form a $d$-dimensional GKP code space
\begin{equation}
	\mathcal{C}_{(\bm{u},\bm{v})} \equiv \{ \ket{\psi} | \hat{S}_X \ket{\psi} = \hat{S}_Z \ket{\psi} = \ket{\psi} \} ,
\end{equation}
if the area $\omega (\bm{u},\bm{v}) = 2\pi d$ and $d$ is an integer. We can equivalently denote a GKP code as $\mathcal{C}_{d,S}$ where $S$ is the normalized basis matrix for the GKP lattice defined as
\begin{equation}
    \label{eq:uv_S}
	\sqrt{2\pi d} S = \begin{pmatrix} u_q & u_p \\ v_q & v_p \end{pmatrix} = \begin{pmatrix} \bm{u}\\ \bm{v} \end{pmatrix} .
\end{equation}
It is easy to see that $S$ is a $2\times 2$ symplectic matrix since $\det (S) = 1$.
Throughout the paper, we will interchangeably choose the notation of either $\mathcal{C}_{(\bm{u},\bm{v})}$ or $\mathcal{C}_{d,S}$ to represent a GKP code based on convenience.

The logical operators for a GKP code $\mathcal{C}_{(\bm{u},\bm{v})}$ are $\hat{X} = \hat{T}(\bm{u}/d)$ and $\hat{Z} = \hat{T}(\bm{v}/d)$ satisfying $\hat{X}\hat{Z} = e^{-2\pi i / d} \hat{Z}\hat{X}$. We choose the eigenstates $\ket{\mu}, \mu=0,...,d-1$ of $\hat{Z}$ as the basis states of the GKP code $\mathcal{C}_{(\bm{u},\bm{v})}$, where
\begin{equation}
    \label{eq:qudit_logical}
	\begin{split}
		\hat{Z} \ket{\mu} =& e^{i \frac{2\pi}{d} \mu} \ket{\mu} \\
		\hat{X} \ket{\mu} =& \ket{\mu + 1 \pmod{d}} .
	\end{split}
\end{equation}
For a square lattice GKP code $\mathcal{C}_{d,S}$ where $S=I_2$ is the $2\times 2$ identity matrix, the basis states in the position space are
\begin{equation}
	\label{eq:GKP_pos_wavefunction}
	\ket{\mu} = \sum_{k=-\infty}^{\infty} \ket{\opq=\sqrt{\frac{2\pi}{d}}(dk+\mu)} , \mu=0,...,d-1 .
\end{equation}
The logical operators for a square lattice GKP code $\mathcal{C}_{d,S}$ are
\begin{equation}
    \label{eq:logical_ops_square_lattice}
    \hat{X} = \exp \left( -i\sqrt{2\pi/d} \opp \right) , \quad \hat{Z} = \exp \left( i\sqrt{2\pi/d} \opq \right) .
\end{equation}
Finally, when $d=1$ the GKP state $\ket{\mu=0}$ is called a qunaught state since it does not carry any quantum information.

\subsection{Explanation of the optimization results}
\label{sec2:explanation}
Inspired by the optimization results, we can construct a concrete example of achieving perfect transmission of a qubit with GKP states. The logical qubit is encoded in the GKP code $\mathcal{C}_{d_1,S_1}$ with $d_1=2$ for mode 1, where the logical basis states are $\ket{\mu_1=0,1}$. The environment state for mode 2 is the qunaught state $\ket{\mu_2=0}$ of the GKP code $\mathcal{C}_{d_2,S_2}$  with $d_2=1$. Here we choose square lattices for both $S_1$ and $S_2$, but later on we will show that $S_1$ and $S_2$ can be arbitrary lattices as long as they satisfy certain lattice matching condition.
Below, we give two intuitive explanations of why logical information of mode 1 is transmitted perfectly from two different perspectives.

\subsubsection{Position space wavefunction}
In the position space, the wavefunctions of the logical basis states are
\begin{equation}
    \label{eq:square_lattice_GKP_d2}
	\begin{split}
		\ket{\mu_1=0} =&  \sum_{k=-\infty}^{\infty} \ket{\opq_1 = 2k \sqrt{\pi}} \\
		\ket{\mu_1=1} =&  \sum_{k=-\infty}^{\infty} \ket{\opq_1 = (2k+1) \sqrt{\pi}} ,
	\end{split}
\end{equation}
and the environment state is
\begin{equation}
    \label{eq:square_lattice_GKP_d1}
	\ket{\mu_2=0} =  \sum_{k=-\infty}^{\infty} \ket{\opq_2 = k \sqrt{2 \pi}} .
\end{equation}
The wavefunctions of $\ket{\mu_1} \ket{\mu_2}$ are shown in upper panel of Fig.~\ref{fig2}(c) where the blue dots represent $\ket{0} \ket{0}$ and orange dots represent $\ket{1} \ket{0}$.

Beam splitter leads to a rotation by an angle $\theta = \arccos \sqrt{\eta}$ in the $(q_1, q_2)$ space since
\begin{equation}
	\opu_\eta \ket{q_1} \ket{q_2} = \ket{\sqrt{\eta} q_1 + \sqrt{1-\eta} q_2} \ket{\sqrt{\eta} q_2 - \sqrt{1-\eta} q_1} .
\end{equation}
Therefore the final states $\ket{\psi_{\mu_1,\mu_2}} \equiv \opu_\eta \ket{\mu_1} \ket{\mu_2}$ can be represented in the lower panel of Fig.~\ref{fig2}(c).
The key observation is that the blue and orange dots overlap with each other when projected to $q_4$, and therefore the marginal state $\oprho_4$ does not carry any logical information.
On the other hand, the blue and orange peaks are well separated for the marginal state $\oprho_3$ leading to perfect transmission of the logical qubit.
Notice that the same picture (Fig.~\ref{fig2}(c)) also holds in the momentum space $(p_1,p_2)$ for the logical qubit in the basis of $\ket{\pm} = (\ket{0} \pm \ket{1}) /\sqrt{2}$, since we are considering square lattice GKP states.

Intuitively, the interference between the two paths $\opq_1 \rightarrow \opq_4$ and $\opq_2 \rightarrow \opq_4$ erases any logical information from the environment, i.e., measuring $\opq_4$ reveals no information about whether $\opq_1$ is logical $0$ or $1$.
Achieving such an erasure requires matching the lattices $S_1,S_2$ of the GKP codes with the rotation angle $\theta$.
Furthermore, the lattice spacings of the marginal distributions reduce by a factor of $\sqrt{3}$ after the beam splitter (Fig.~\ref{fig2}(c)), indicating that the output state is embedded in a larger dimensional GKP code, which we will derive later on.

\subsubsection{Characteristic function}
Characteristic function is a useful tool for understanding Gaussian channels, since the state transformation is equivalent to multiplication of the characteristic functions.
The characteristic function $\chi_{\oprho} (\bm{\alpha})$ of a state $\oprho$ is defined as the expectation value of a displacement operator:
\begin{equation}
	\chi_{\oprho} (\bm{\alpha}) = \tr \left[ \oprho \hat{T} (\bm{\alpha}) \right] .
\end{equation}
When sending $\oprho_1 \otimes \oprho_2$ into a beam splitter, the characteristic functions of the marginal states $\oprho_3$ and $\oprho_4$ (Eq.~(\ref{eq:marginal_states})) are given by
\begin{equation}
	\label{eq:output_chi}
	\begin{split}
		\chi_3 (\bm{\alpha}) =& \chi_1 (\sqrt{\eta} \bm{\alpha}) \chi_2 (\sqrt{1-\eta} \bm{\alpha}) \\
		\chi_4 (\bm{\alpha}) =& \chi_1 (-\sqrt{1-\eta} \bm{\alpha}) \chi_2 (\sqrt{\eta} \bm{\alpha}) ,
	\end{split}
\end{equation}
where $\chi_k (\bm{\alpha})$ is the characteristic function of $\oprho_k,k=1\sim 4$.

The characteristic function of a GKP state is a periodic 2D lattice, which can be represented by its unit cell.
In Fig.~\ref{fig2}(d), we plot $\chi_1(\bm{\alpha})$ and $\chi_2(\bm{\alpha})$ before the beam splitter where non-zero values are only taken on the black and red lattice points.
The black dots correspond to the stabilizer measurements and therefore the values of the characteristic function at the black dots are always 1, which is the trace of the logical qubit, $\tr [ \oprho ]$.
The red dots correspond to the logical measurements and the values of the characteristic function at the red dots are the expectation values, $\tr[\oprho \hat{X}]$, $\tr[\oprho \hat{Y}]$, $\tr[\oprho \hat{Z}]$, respectively.
For example, if $\oprho_1$ is $\ket{\mu_1=0}$ or $\ket{\mu_1=1}$, the $Z$ dot would be $1$ or $-1$ while the $X,Y$ dots would be 0.

Our goal is to ``hide'' the values of characteristic function at these special red points from the environment, since they carry logical information.
After the beam splitter, the characteristic functions of the output states are shown in Fig.~\ref{fig2}(d). From Eq.~(\ref{eq:output_chi}), we can view $\chi_2 (\sqrt{1-\eta} \bm{\alpha})$ and $\chi_2 (\sqrt{\eta} \bm{\alpha})$ as filters of the logical information stored in $\chi_1(\bm{\alpha})$. Because the filtering, the red dots are preserved in $\chi_3(\bm{\alpha})$ but not in $\chi_4(\bm{\alpha})$. As a result, the logical information in $\oprho_1$ is erased from $\oprho_4$ but preserved in $\oprho_3$.

\section{General theory}
\label{sec3}
In this section, we first give a general condition for achieving perfect transmission for an environment-assisted quantum channel. We then construct the input encoding and environment state for the beam splitter channel and characterize the entanglement of the output states. Finally we provide a few examples corresponding to various applications.

\subsection{Perfect transmission condition}
\label{sec:code_construction}
Consider an environment-assisted quantum channel $\mathcal{E}_{\hat{\sigma}}$ (Eq.~(\ref{eq:env_assisted_channel})) associated with a unitary transformation $\opu$ acting on the system $\mathcal{H}_S$ and its environment $\mathcal{H}_E$.
We can encode quantum information in a logical subspace $\mathcal{C} \in \mathcal{H}_S$ of the system with basis states $\{\ket{\psi_k}\}$.
The encoded quantum information is preserved under $\opu$, i.e., noises from the environment are correctable, if and only if
\begin{equation}
	\label{eq:QEC_conditions}
	\tr_S \left[ \opu(\ket{\psi_i} \bra{\psi_j} \otimes \ket{\psi_E} \bra{\psi_E}) \opud \right] = \delta_{ij} \oprho_E, \qquad \forall i,j ,
\end{equation}
where $\hat{\sigma}=\ket{\psi_E} \bra{\psi_E}$ and $\oprho_E$ are the environment states before and after $\opu$. Intuitively, this means the environment state after $\opu$ does not contain any logical information, thus enabling perfect transmission.
We can prove Eq.~(\ref{eq:QEC_conditions}) from the standard Knill-Laflamme conditions~\cite{nielsen2012} in Appendix~\ref{sec:SI_proof_1} and the conclusion only holds if the environment state $\hat{\sigma}$ is pure.

\begin{figure*}[t]
	\centering
	\includegraphics[width=\textwidth]{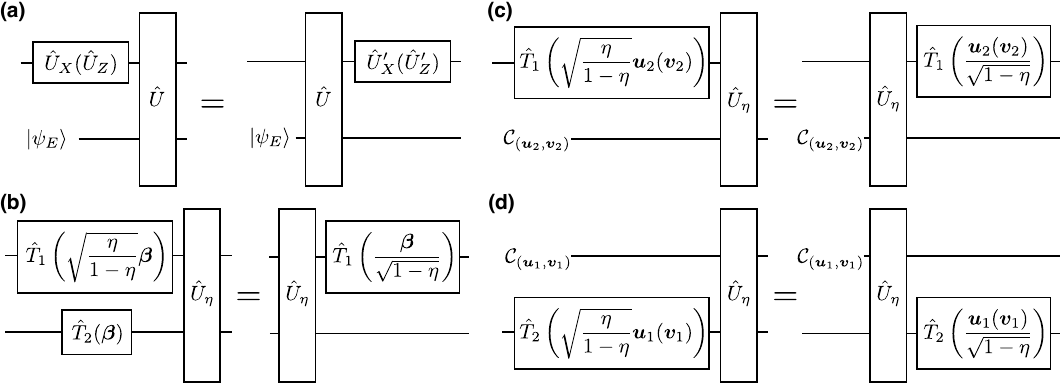}
	\caption{(a) Unitary operators $\opu_X$ and $\opu_Z$ acting on the system before $\opu$ does not change the environment state after $\opu$. (b) Commutation relation between displacement operators and the beam splitter unitary $\opu_\eta$. (c) Circuit equality when the input state of mode 2 is any GKP state from $\mathcal{C}_{(\bm{u}_2, \bm{v}_2)}$, derived by choosing $\bm{\beta}=\bm{u}_2$ or $\bm{\beta}=\bm{v}_2$ in (b). (d) Circuit equality when the input state of mode 1 is any GKP state from $\mathcal{C}_{(\bm{u}_1, \bm{v}_1)}$.}
	\label{fig3}
\end{figure*}

\subsubsection{Encoding from operators}
Instead of directly solving the logical subspace $\mathcal{C}$ from Eq.~(\ref{eq:QEC_conditions}), we will construct $\mathcal{C}$ from certain operators with nice properties.
The construction is based on finding unitary operators $\opu_X (\opu_Z)$ and $\opu_X' (\opu_Z')$ that only act on the system and satisfy the equality in Fig.~\ref{fig3}(a), i.e., for any system state $\ket{\psi}$
\begin{equation}
    \begin{split}
        \opu \opu_X \ket{\psi} \ket{\psi_E} =& \opu_X' \opu \ket{\psi} \ket{\psi_E} \\
        \opu \opu_Z \ket{\psi} \ket{\psi_E} =& \opu_Z' \opu \ket{\psi} \ket{\psi_E} .
    \end{split}
\end{equation}
In other words, the final environment state is the same regardless of whether $\opu_X,\opu_Z$ act on the system before $\opu$.
Furthermore, we assume the commutation relation
\begin{equation}
	\label{eq:XZ_commutator}
	\opu_X \opu_Z  = e^{-i\phi} \opu_Z \opu_X ,
\end{equation}
where $\phi \geq 0$ since otherwise we can swap the labels $X$ and $Z$.

Given $\opu_X,\opu_Z$, the logical basis states $\{ \ket{\psi_k} \}$ satisfying Eq.~(\ref{eq:QEC_conditions}) can be constructed by acting $\opu_X$ on the eigenstates of $\opu_Z$.
More specifically, starting from any eigenstate of $\opu_Z$ with eigenvalue $e^{i\phi_0}$, we define the logical subspace as
\begin{equation}
    \label{eq:general_code}
    \mathcal{C} = \text{span}\left \{\ket{\psi_k} = (\opu_X)^k \ket{\psi_0}, k=0,...,d-1 \right \} .
\end{equation}
From Eq.~(\ref{eq:XZ_commutator}), we know that $\ket{\psi_k}$ is also an eigenstate of $\opu_Z$ with eigenvalue $e^{i\phi_k}$, where $\phi_k = \phi_0 + k \phi$.
If the eigenvalues $e^{i \phi_0},...,e^{i \phi_{d-1}}$ are all different, i.e., $\phi_i \equiv \phi_j \pmod{2\pi}$ if and only if $i=j$, then $\{ \ket{\psi_k} \}$ are mutually orthogonal and span a $d$-dimensional subspace $\mathcal{C}$.
It turns out that:

\begin{lemma}
    The subspace $\mathcal{C}$ constructed in Eq.~(\ref{eq:general_code}) satisfies the error correction condition Eq.~(\ref{eq:QEC_conditions}) and therefore is preserved under $\opu$.
\end{lemma}

\begin{proof}
    For $i \neq j$, we have
    \begin{equation}
    	\begin{split}
    		& \tr_S \left[ \opu ( \opu_Z \ket{\psi_i} \bra{\psi_j} \opud_Z \otimes \ket{\psi_E} \bra{\psi_E}) \opud \right] \\
    		=& \tr_S \left[ \opu_Z' \opu (\ket{\psi_i} \bra{\psi_j} \otimes \ket{\psi_E} \bra{\psi_E}) \opud \opu_Z^{\prime \dagger} \right] \\
    		=& \tr_S \left[ \opu (\ket{\psi_i} \bra{\psi_j} \otimes \ket{\psi_E} \bra{\psi_E}) \opud \right] \\
    		=& e^{i (\phi_i-\phi_j)} \tr_S \left[ \opu (\ket{\psi_i} \bra{\psi_j} \otimes \ket{\psi_E} \bra{\psi_E}) \opud \right] = 0.
    	\end{split}
    \end{equation}
    For $i=j<d-1$, we have
    \begin{equation}
    	\begin{split}
    		& \tr_S \left[ \opu ( \ket{\psi_i} \bra{\psi_i} \otimes \ket{\psi_E} \bra{\psi_E}) \opud \right] \\
            =& \tr_S \left[ \opu_X' \opu ( \ket{\psi_i} \bra{\psi_i} \otimes \ket{\psi_E} \bra{\psi_E}) \opud \opu_X^{\prime \dagger} \right] \\
    		=& \tr_S \left[ \opu ( \opu_X \ket{\psi_i} \bra{\psi_i} \opud_X \otimes \ket{\psi_E} \bra{\psi_E}) \opud \right] \\
    		=& \tr_S \left[ \opu (\ket{\psi_{i + 1}} \bra{\psi_{i + 1}} \otimes \ket{\psi_E} \bra{\psi_E}) \opud \right] \equiv \oprho_E .
    	\end{split}
    \end{equation}
\end{proof}

If $\phi/2\pi$ is irrational in Eq.~(\ref{eq:XZ_commutator}), the dimension $d$ can be made arbitrarily large since $(i-j) \phi \equiv 0 \pmod{2\pi}$ is only possible if $i=j$. In this case, the environment-assisted quantum channel $\mathcal{E}_{\hat{\sigma}}$ can transmit an infinite dimensional subspace perfectly and therefore its quantum capacity is infinite without energy constraint. On the other hand, if $\phi/2\pi=m/n$ is rational where integers $m,n$ are coprime, i.e., $\gcd(m,n)=1$, it is possible to transmit a $n$-dimensional subspace perfectly.

\subsubsection{GKP encoding from displacement operators}
When $\opu$ is a two-mode Gaussian unitary such as beam splitter and two-mode squeezing, it turns out that we can choose $\ket{\psi_E}$ such that $\opu_X=\hat{T}(\bm{\tilde{u}})$ and $\opu_Z=\hat{T}(\bm{\tilde{v}})$ are two displacement operators with $\phi = \omega (\bm{\tilde{u}}, \bm{\tilde{v}})$.

There are infinite constructions of $\mathcal{C}$ by choosing $\ket{\psi_0}$ as different eigenstates of $\opu_Z$.
One possible choice for $\{\ket{\psi_k}\}$ are the quadrature eigenstates.
As a concrete example, we may choose the following logical basis states for $\opa_1$ instead of Eq.~(\ref{eq:square_lattice_GKP_d2}):
\begin{equation}
    \label{eq:other_encodings}
	\begin{split}
		\ket{\psi_0} =&  \ket{\opq_1 = 0} \\
		\ket{\psi_1} =&  \ket{\opq_1 = \sqrt{\pi}} ,
	\end{split}
\end{equation}
which also leads to perfect transmission of one qubit.
The erasure of the logical information from the environment can be intuitively understood from Fig.~\ref{fig2}(c), where instead of having an infinite two-dimensional grid, we now have a $2\times \infty$ grid.
In practice, we can approximate Eq.~(\ref{eq:other_encodings}) with squeezed cat states which may be easier to implement than GKP states in Eq.~(\ref{eq:square_lattice_GKP_d2}).

For the rest of the paper, we select $\ket{\psi_0}$ as a GKP state and $\ket{\psi_k}$, obtained by displacing $\ket{\psi_0}$, is also a GKP state.
The choice of GKP encodings offers nice properties, such as the ability of simultaneous bidirectional quantum communication through the beam splitter, which will be explained later.

For irrational $\phi/2\pi$, the resulting subspace $\mathcal{C}$ from Eq.~(\ref{eq:general_code}) is not the usual GKP code, although its basis states are GKP states. For example, $\mathcal{C}$ may look like
\begin{equation}
    \label{eq:irrational_GKP}
	\begin{split}
		\ket{\psi_0} =&  \sum_{k=-\infty}^{\infty} \ket{\opq = 2k \sqrt{\pi}} \\
		\ket{\psi_1} =&  \sum_{k=-\infty}^{\infty} \ket{\opq = (2k+\epsilon) \sqrt{\pi}} ,
	\end{split}
\end{equation}
where $\epsilon$ is irrational. In this case, the subspace $\mathcal{C} = \text{span} \{ \ket{\psi_0}, \ket{\psi_1} \}$ is only invariant under displacements along $\opq$ by multiples of $2\sqrt{\pi}$, yet it lacks invariance under any displacement along $\opp$. In comparison, conventional GKP codes such as Eq.~(\ref{eq:square_lattice_GKP_d2}) have two types of stabilizers along both $\opq$ and $\opp$.

For rational $\phi/2\pi$, $\mathcal{C}$ is a finite dimensional GKP code. Let
\begin{equation}
	\phi = \omega (\bm{\tilde{u}}, \bm{\tilde{v}}) = 2\pi\frac{m}{kd} ,
\end{equation}
where integers $m,k,d$ satisfy $\gcd(m,kd)=1$, we can construct $\mathcal{C}$ as a $d$-dimensional GKP code $\mathcal{C}_{(\bm{u},\bm{v})}$ with stabilizers $\hat{S}_X=\hat{T}(\bm{u})$ and $\hat{S}_Z=\hat{T}(\bm{v})$, where
\begin{equation}
	\label{eq:construction}
	\bm{u} = \frac{d k_1}{m_1} \bm{\tilde{u}} , \qquad \bm{v} = \frac{d k_2}{m_2} \bm{\tilde{v}} ,
\end{equation}
and integers $m_1,m_2,k_1,k_2$ satisfy $m = m_1 m_2$ and $k=k_1 k_2$ (see Appendix~\ref{sec:SI_proof_2}).

\subsection{Encoding and environment state for the beam splitter channel}
\label{sec3:main_result}
Here we derive the encoding and environment state to achieve perfection transmission through a beam splitter.
The idea is to first find the environment state $\ket{\psi_E}$ such that the equality in Fig.~\ref{fig3}(a) holds, and $\opu_X,\opu_Z$ turn out to be displacement operators.
After that we can apply Eq.~(\ref{eq:construction}) to construct the encoding.
Similar constructions can also be derived for other two-mode Gaussian unitaries (see Appendix~\ref{sec:SI_other_transformation}).

Displacement operators have the following commutation relation with the beam splitter unitary $\opu_\eta$:
\begin{equation}
	\begin{split}
		& \opu_\eta \hat{T}_1(\bm{\alpha}) \hat{T}_2(\bm{\beta}) \\
		=& \hat{T}_1(\sqrt{\eta}\bm{\alpha} + \sqrt{1-\eta} \bm{\beta}) \hat{T}_2 (\sqrt{\eta}\bm{\beta} - \sqrt{1-\eta} \bm{\alpha}) \opu_\eta ,
	\end{split}
\end{equation}
where $\hat{T}_k,k=1,2$ is the displacement operator of mode $k$ and $\bm{\alpha},\bm{\beta}$ can be arbitrary vectors.
This leads to the equality in Fig.~\ref{fig3}(b) which holds for any input states and any $\bm{\beta}$.

The key observation is that we can choose $\ket{\psi_E}$, the input state of mode 2, as a GKP state to absorb the operator $\hat{T}_2 (\bm{\beta})$ in Fig.~\ref{fig3}(b), since GKP states are invariant under certain displacements.
More concretely, we choose the input state of mode 2 from a $d_2$-dimensional GKP code $\mathcal{C}_{(\bm{u}_2, \bm{v}_2)}$ with $\omega(\bm{u}_2, \bm{v}_2) = 2\pi d_2$. This leads to the equality in Fig.~\ref{fig3}(c) which is the same as Fig.~\ref{fig3}(a) with
\begin{equation}
    \begin{split}
        \opu_X =& \hat{T}_1\left( \sqrt{\frac{\eta}{1-\eta}} \bm{u}_2 \right) \equiv \hat{T}_1 (\bm{\tilde{u}}_1) \\
        \opu_Z =& \hat{T}_1\left( \sqrt{\frac{\eta}{1-\eta}} \bm{v}_2 \right) \equiv \hat{T}_1 (\bm{\tilde{v}}_1) ,
    \end{split}
\end{equation}
and the phase is
\begin{equation}
    \phi = \omega (\bm{\tilde{u}}_1, \bm{\tilde{v}}_1) = \frac{\eta}{1-\eta} \omega(\bm{u}_2, \bm{v}_2) = \frac{\eta}{1-\eta} 2\pi d_2 .
\end{equation}
For irrational $\eta$, $\phi/2\pi$ is also irrational which allows perfect transmission of an infinite dimensional subspace through $\mathcal{E}_1$.
Therefore we will mostly focus on rational $\eta$ where the logical subspace is a GKP code.

Perfect transmission of a $d_1$-dimensional GKP code $\mathcal{C}_{(\bm{u}_1, \bm{v}_1)}$ for mode 1 requires
\begin{equation}
    \omega (\bm{\tilde{u}}_1, \bm{\tilde{v}}_1) = 2\pi \frac{m}{k d_1} ,
\end{equation}
which leads to $\eta=m/n$ with $n=m+k d_1 d_2 = m_1 m_2 + k_1 k_2 d_1 d_2$.
The explicit construction from Eq.~(\ref{eq:construction}) gives the lattice matching condition
\begin{equation}
	\label{eq:lattice_matching_uv}
	\begin{split}
		\bm{u}_1 =& \frac{d_1 k_1}{m_1} \bm{\tilde{u}}_1 = \frac{d_1 k_1}{m_1} \sqrt{\frac{\eta}{1-\eta}} \bm{u}_2 = \sqrt{\frac{d_1}{d_2}} \sqrt{\frac{k_1 m_2}{k_2 m_1}} \bm{u}_2 \\
		\bm{v}_1 =& \frac{d_1 k_2}{m_2} \bm{\tilde{v}}_1 = \frac{d_1 k_2}{m_2} \sqrt{\frac{\eta}{1-\eta}} \bm{v}_2 = \sqrt{\frac{d_1}{d_2}} \sqrt{\frac{k_2 m_1}{k_1 m_2}} \bm{v}_2 .
	\end{split}
\end{equation}

Interestingly, with Eq.~(\ref{eq:lattice_matching_uv}) it is possible to simultaneously achieve perfect transmission of $\mathcal{C}_{(\bm{u}_2, \bm{v}_2)}$ for mode 2 when treating mode 1 as the environment (Fig.~\ref{fig3}(d)), where
\begin{equation}
    \begin{split}
        \opu_X =& \hat{T}_2\left( \sqrt{\frac{\eta}{1-\eta}} \bm{u}_1 \right) \equiv \hat{T}_2 (\bm{\tilde{u}}_2) \\
        \opu_Z =& \hat{T}_2\left( \sqrt{\frac{\eta}{1-\eta}} \bm{v}_1 \right) \equiv \hat{T}_2 (\bm{\tilde{v}}_2) ,
    \end{split}
\end{equation}
Notice that perfect transmission of $\mathcal{C}_{(\bm{u}_1, \bm{v}_1)}$ only requires $\gcd(m,kd_1)=1$, while adding the extra requirement of $\gcd(m,d_2)=1$ leads to simultaneous perfect transmission of $\mathcal{C}_{(\bm{u}_2, \bm{v}_2)}$ since
\begin{equation}
    \phi = \omega (\bm{\tilde{u}}_2, \bm{\tilde{v}}_2) = \frac{\eta}{1-\eta} \omega(\bm{u}_1, \bm{v}_1) = 2\pi \frac{m}{k d_2} .
\end{equation}
We can verify the perfect transmission of $\mathcal{C}_{(\bm{u}_2, \bm{v}_2)}$ by rewriting Eq.~(\ref{eq:lattice_matching_uv}) as
\begin{equation}
	\label{eq:lattice_matching_uv_2}
	\begin{split}
		\bm{u}_2 =& \sqrt{\frac{d_2}{d_1}} \sqrt{\frac{k_2 m_1}{k_1 m_2}} \bm{u}_1 = \frac{d_2 k_2}{m_2} \sqrt{\frac{\eta}{1-\eta}} \bm{u}_1 = \frac{d_2 k_2}{m_2} \bm{\tilde{u}}_2 \\
		\bm{v}_2 =& \sqrt{\frac{d_2}{d_1}} \sqrt{\frac{k_1 m_2}{k_2 m_1}} \bm{v}_1 = \frac{d_2 k_1}{m_1} \sqrt{\frac{\eta}{1-\eta}} \bm{v}_1 = \frac{d_2 k_1}{m_1} \bm{\tilde{v}}_2 ,
	\end{split}
\end{equation}
which also satisfies the construction Eq.~(\ref{eq:construction}).

\begin{figure*}[t]
	\centering
    \begin{tikzpicture}[
    node distance=0.8cm,
    node/.style={
      rectangle,
      rounded corners,
      inner sep=2mm,
      text centered,
      draw=black}
    ]
    \node[anchor=north west,inner sep=0] (main) at (0,0) {\includegraphics[width=\textwidth]{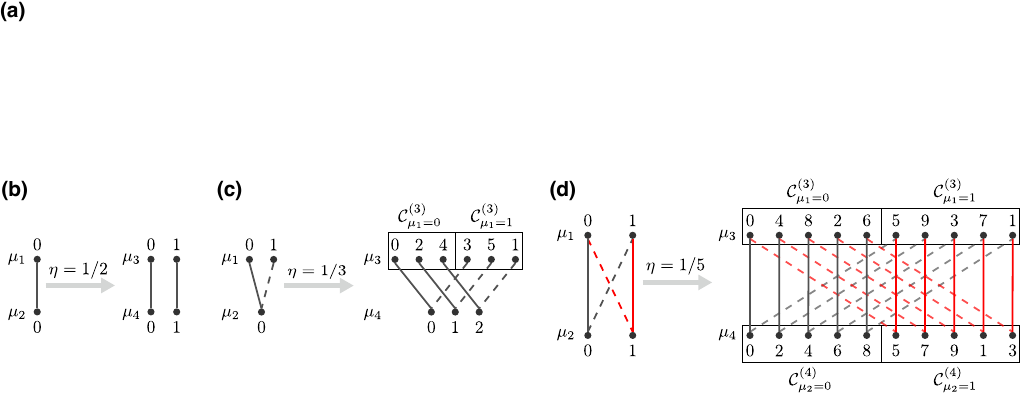}};
    \begin{scope}
        \node (step1) [anchor=north west,node] at (0.2,-0.8) {$\begin{aligned} & \eta = \frac{m}{n} \\ & \gcd(m,n)=1 \end{aligned}$};
        \node (step2) [node, right=of step1] {$\begin{aligned} & n = m + k d_1 d_2 \\ & m = m_1 m_2 \\ & k = k_1 k_2 \end{aligned}$};
        \node (step3) [node, right=of step2] {$\begin{aligned}
        & \mathcal{C}_{d_1,S_1}, \mathcal{C}_{d_2,S_2} \\
        & S_2 = \begin{pmatrix} \sqrt{\frac{k_2 m_1}{k_1 m_2}} & 0 \\ 0 & \sqrt{\frac{k_1 m_2}{k_2 m_1}} \end{pmatrix} S_1
        \end{aligned}$};
        \node (step4) [node, right=of step3] {$\begin{aligned}
        & \mathcal{C}_{d_3,S_3}, \mathcal{C}_{d_4,S_4} \text{~with~} d_3 = d_1 n, d_4 = d_2 n \\
        & S_3 = \begin{pmatrix} \sqrt{m_1/m_2} & 0 \\ 0 & \sqrt{m_2/m_1} \end{pmatrix} S_1  \\
        & S_4 = \begin{pmatrix} \sqrt{m_2/m_1} & 0 \\ 0 & \sqrt{m_1/m_2} \end{pmatrix} S_2 
        \end{aligned}$};
        \draw[->, line width=1pt] (step1) -- (step2);
        \draw[->, line width=1pt] (step2) -- (step3);
        \draw[->, line width=1pt] (step3) -- (step4);
    \end{scope}
    \end{tikzpicture}
    
	\caption{(a) Flow diagram illustrating the selection of the input GKP encodings $\mathcal{C}_{d_1,S_1}, \mathcal{C}_{d_2,S_2}$ for a given $\eta$. (b) Generating a GKP Bell state at $\eta=1/2$, where the input code dimensions are $(d_1,d_2)=(1,1)$ and the output code dimensions are $(d_3,d_4)=(2,2)$. (c) Unidirectional quantum transduction at $\eta=1/3$ with $(d_1,d_2,d_3,d_4)=(2,1,6,3)$. (d) Duplex quantum transduction at $\eta=1/5$ with $(d_1,d_2,d_3,d_4)=(2,2,10,10)$.}
	\label{fig4}
\end{figure*}

Our main result (Fig.~\ref{fig_main_result} and Fig.~\ref{fig4}(a)) is:
\begin{theorem}
    \label{main_theorem}
    For a beam splitter $\opu_\eta$ with rational transmissivity $\eta = m/n$ where $\gcd(m,n)=1$ and $n=m+k d_1 d_2$, a $d_1$-dimensional GKP code $\mathcal{C}_{d_1,S_1}$ of mode 1 and a $d_2$-dimensional GKP code $\mathcal{C}_{d_2,S_2}$ of mode 2 can be perfectly transmitted simultaneously. The prefect transmission can be achieved with the lattice matching condition (c.f., Eq.~(\ref{eq:uv_S}) and Eq.~(\ref{eq:lattice_matching_uv}))
    \begin{equation}
    	\label{eq:lattice_matching}
    	S_2 = \begin{pmatrix} \sqrt{\frac{k_2 m_1}{k_1 m_2}} & 0 \\ 0 & \sqrt{\frac{k_1 m_2}{k_2 m_1}} \end{pmatrix} S_1 ,
    \end{equation}
    where $m=m_1 m_2, k=k_1 k_2$ and $m_1,m_2,k_1,k_2$ are positive integers.
\end{theorem}

Non-trivial quantum communication is possible with any rational transmissivities, except for $\eta=(n-1)/n, n \geq 2$ which only works for the trivial case $d_1=d_2=1$. Transmission of a $n-m$ dimensional GKP code through a beam splitter is always possible by choosing $k=1$. Furthermore, for a given $\eta$ there can be multiple choices of the code dimensions. For example, we can have $(d_1,d_2)=(2,2),(4,1),(2,1)$ for $\eta=1/5$.

\subsection{Output states from the beam splitter channel}
\label{sec3:output_states}
Here we summarize our results which characterize the output states from the beam splitter channel. See Appendix~\ref{sec:SI_output_states} for detailed derivations.

When the input states of the beam splitter are chosen from $\mathcal{C}_{d_1,S_1} \otimes \mathcal{C}_{d_2,S_2}$ where $S_1$ and $S_2$ satisfy the lattice matching condition Eq.~(\ref{eq:lattice_matching}), the output states are embedded in a larger dimensional GKP code $\mathcal{C}_{d_3,S_3} \otimes \mathcal{C}_{d_4,S_4}$, where $d_3 = d_1 n, d_4 = d_2 n$ and
\begin{equation}
    \label{eq:output_lattice}
	\begin{split}
		S_3 =& \begin{pmatrix} \sqrt{\frac{m_1}{m_2}} & 0 \\ 0 & \sqrt{\frac{m_2}{m_1}} \end{pmatrix} S_1 = \begin{pmatrix} \sqrt{\frac{k_1}{k_2}} & 0 \\ 0 & \sqrt{\frac{k_2}{k_1}} \end{pmatrix} S_2 \\ 
		S_4 =& \begin{pmatrix} \sqrt{\frac{k_2}{k_1}} & 0 \\ 0 & \sqrt{\frac{k_1}{k_2}} \end{pmatrix} S_1 = \begin{pmatrix} \sqrt{\frac{m_2}{m_1}} & 0 \\ 0 & \sqrt{\frac{m_1}{m_2}} \end{pmatrix} S_2 .
	\end{split}
\end{equation}
More precisely, embedding means that the marginal states in Eq.~(\ref{eq:marginal_states}) satisfy $\oprho_3 \in \mathcal{C}_{d_3,S_3}$ and $\oprho_4 \in \mathcal{C}_{d_4,S_4}$, while the joint output state is entangled between the two modes (Fig.~\ref{fig_main_result} and Fig.~\ref{fig4}(a)).

Let $\{ \ket{\mu_l}, \mu_l=0,...,d_l-1 \}$ be the basis states of $\mathcal{C}_{d_l,S_l},l=1\sim 4$, the input state $\ket{\mu_1} \ket{\mu_2}$ leads to the output state
\begin{equation}
    \ket{\psi_{\mu_1,\mu_2}} \equiv \opu_\eta \ket{\mu_1} \ket{\mu_2} .
\end{equation}
The larger dimensional GKP codes at the output can be decomposed as a direct sum
\begin{equation}
    \mathcal{C}_{d_3,S_3} = \oplus_{\mu_1=0}^{d_1-1} \mathcal{C}^{(3)}_{\mu_1} , \qquad \mathcal{C}_{d_4,S_4} = \oplus_{\mu_2=0}^{d_2-1} \mathcal{C}^{(4)}_{\mu_2} ,
\end{equation}
where $|\mathcal{C}^{(3)}_{\mu_1}| = |\mathcal{C}^{(4)}_{\mu_2}| = n$ and
\begin{equation}
    \label{eq:direct_sum}
	\begin{split}
		\mathcal{C}_{\mu_1}^{(3)} \equiv & \{ \ket{\mu_3} \mid \mu_3 \equiv \mu_1 m_2 \pmod{d_1} \} \\
		=& \{ \ket{\mu_1 m_2 + j d_1 \pmod{d_3}} ,j=0,...,n-1 \} , \\
		\mathcal{C}_{\mu_2}^{(4)} \equiv & \{ \ket{\mu_4} \mid \mu_4 \equiv \mu_2 m_1 \pmod{d_2} \} \\
		=& \{ \ket{\mu_2 m_1 + j d_2 \pmod{d_4}} ,j=0,...,n-1 \} .
	\end{split}
\end{equation}
The output state $\ket{\psi_{\mu_1,\mu_2}}$ is one of the maximally entangled states in $\mathcal{C}^{(3)}_{\mu_1} \otimes \mathcal{C}^{(4)}_{\mu_2}$, and different $\ket{\psi_{\mu_1,\mu_2}}$ belongs to different subspaces.
We can explicitly represent $\ket{\psi_{\mu_1,\mu_2}}$ in the basis of $\ket{\mu_3} \ket{\mu_4}$ as
\begin{equation}
    \label{eq:output_states}
	\ket{\psi_{\mu_1,\mu_2}} = \frac{1}{\sqrt{n}} \sum_{j=0}^{n-1} \ket{\mu_1 \alpha_1 n + j k_2 d_2 d_1} \ket{\mu_2 \alpha_2 n + j m_1 d_2} ,
\end{equation}
where integer $\alpha_l$ satisfies $m_l \alpha_l \equiv 1 \pmod{d_l},l=1,2$. We have omitted the module $d_3$ and $d_4$ on the right hand side for notational simplicity.

The maximal entanglement provides another way to explain the simultaneous perfect transmission, since the marginal states $\oprho_3$ and $\oprho_4$ carry no information about $\mu_2$ and $\mu_1$.
Furthermore, it is possible to perform a subsystem decomposition~\cite{pantaleoni2020,shaw2022} of $\mathcal{C}_{d_3,S_3} (\mathcal{C}_{d_4,S_4})$ for each output mode $\opa_3 (\opa_4)$ into a logical subsystem and a gauge subsystem.
The two gauge subsystems become maximally entangled, while the two logical subsystems are not entangled.
The input GKP codes $\mathcal{C}_{d_1,S_1}$ and $\mathcal{C}_{d_2,S_2}$ are mapped to the logical subsystems of $\opa_3$ and $\opa_4$ respectively by the beam splitter.

Finally, we can study the transformation of the logical operators. Before the beam splitter, the logical operators are
\begin{equation}
    \label{eq:input_logical_ops}
    \begin{split}
        \hat{X}_1 =& \hat{T}_1 (\bm{u}_1/d_1), \qquad \hat{Z}_1 = \hat{T}_1 (\bm{v}_1/d_1) , \\
        \hat{X}_2 =& \hat{T}_2 (\bm{u}_2/d_2), \qquad \hat{Z}_2 = \hat{T}_2 (\bm{v}_2/d_2) ,
    \end{split}
\end{equation}
where $\bm{u}_l,\bm{v}_l$ are the lattice vectors of $\mathcal{C}_{d_l,S_l}$.
After the beam splitter, the output logical operators are $\hat{X}_1', \hat{Z}_1', \hat{X}_2', \hat{Z}_2'$ where $\hat{O}' \equiv \opu_\eta \hat{O} \opud_\eta$, which in general act on both output modes.
However, simultaneous perfect transmission guarantees that we can find equivalent logical operators of the output states which only act on one output mode.
These single mode logical operators are
\begin{equation}
    \label{eq:output_logical_ops_single}
	\begin{split}
		\tilde{X}_1 =& \hat{T}_1 (\alpha_1 \bm{u}_3/d_1) \Leftrightarrow \hat{X}_1' \\
        \tilde{Z}_1 =& \hat{T}_1 (\beta_1 \bm{v}_3/d_1) \Leftrightarrow \hat{Z}_1' \\
		\tilde{X}_2 =& \hat{T}_2 (\alpha_2 \bm{u}_4/d_2) \Leftrightarrow \hat{X}_2' \\
		\tilde{Z}_2 =& \hat{T}_2 (\beta_2 \bm{v}_4/d_2) \Leftrightarrow \hat{Z}_2' ,
	\end{split}
\end{equation}
where integers $\beta_1$ and $\beta_2$ satisfy $\beta_1 m_2 \equiv 1 \pmod{d_1}$ and $\beta_2 m_1 \equiv 1 \pmod{d_2}$, and $\Leftrightarrow $ means equivalent when acting on the output states.
Physically, measuring $\tilde{X}_1 (\tilde{X}_2)$ and $\tilde{Z}_1 (\tilde{Z}_2)$ on the marginal output state $\oprho_3 (\oprho_4)$ is equivalent to measuring $\hat{X}_1 (\hat{X}_2)$ and $\hat{Z}_1 (\hat{Z}_2)$ on the input state $\oprho_1 (\oprho_2)$, and thus provides a way to perform logical state tomography of the output states.

\subsection{Examples}
\label{sec3:examples}
In this section, we present three illustrative examples for different applications, and characterize their output states as well as the single-mode logical operators associated with these states. For all examples, we have $k_1=k_2=m_1=m_2=1$ leading to $S_1=S_2=S_3=S_4 \equiv S$. Therefore, the lattices of the GKP codes for both input and output modes are identical, where $S$ can be arbitrary and the code dimensions may differ.

\subsubsection{Entanglement generation ($\eta=1/2$)}
For $\eta=1/2$, we have $n=2$ and $d_1=d_2=1$, which leads to $d_3=d_4=2$. The input state $\ket{0}\ket{0}$ is a product state of GKP qunaught states and the output state (Eq.~(\ref{eq:output_states})) is a GKP Bell state (Fig.~\ref{fig4}(b))
\begin{equation}
    \ket{\psi_{0,0}} = \frac{1}{\sqrt{2}} (\ket{0}\ket{0} + \ket{1}\ket{1}) .
\end{equation}
Generating a GKP Bell state by sending a product state of two GKP qunaught states into a $50/50$ beam splitter has been discovered previously~\cite{walshe2020} for rectangle lattices, and here we generalize this result to arbitrary lattices $S$.

\subsubsection{Unidirectional quantum transduction ($\eta=1/3$)}
For $\eta=1/3$, we have $n=3$ and $(d_1,d_2) = (2,1)$, which leads to $(d_3,d_4)=(6,3)$. The output states (Eq.~(\ref{eq:output_states})) are (Fig.~\ref{fig4}(c))
\begin{equation}
    \begin{split}
        \ket{\psi_{0,0}} =& \frac{1}{\sqrt{3}} \left(  \ket{0}\ket{0} + \ket{2}\ket{1} + \ket{4}\ket{2} \right) \\
        \ket{\psi_{1,0}} =& \frac{1}{\sqrt{3}} \left( \ket{3}\ket{0} + \ket{5}\ket{1} + \ket{1}\ket{2} \right) ,
    \end{split}
\end{equation}
which enables perfect unidirectional quantum transduction from $\opa_1$ to $\opa_3$, as we have already obtained from the numerical optimization in Sec.~\ref{sec2}.
Furthermore, the code dimensions $(d_3,d_4)$ increase by a factor of $n=3$ compared to $(d_1,d_2)$, which explains why the lattice spacings of the marginal distributions reduce by a factor of $\sqrt{3}$ in Fig.~\ref{fig2}(c) for a square lattice $S$.

We can choose $\alpha_1=\alpha_2=\beta_1=\beta_2 = 1$ and the single mode logical operators (Eq.~(\ref{eq:output_logical_ops_single})) scale the displacement vectors of the input logical operators by a factor of $\sqrt{n}$, i.e., $\tilde{X}_1 = (\hat{X}_1)^{\sqrt{3}}, \tilde{Z}_1 = (\hat{Z}_1)^{\sqrt{3}}, \tilde{X}_2 = (\hat{X}_2)^{\sqrt{3}}, \tilde{Z}_2 = (\hat{Z}_2)^{\sqrt{3}}$.
For a square lattice $S$, we have
\begin{equation}
    \begin{split}
        \tilde{X}_1 =& \exp \left( -i\sqrt{3\pi} \opp_1 \right) , \quad \tilde{Z}_1 = \exp \left( i\sqrt{3\pi} \opq_1 \right) , \\
        \tilde{X}_2 =& \exp \left( -i\sqrt{6\pi} \opp_2 \right) , \quad \tilde{Z}_2 = \exp \left( i\sqrt{6\pi} \opq_2 \right) ,
    \end{split}
\end{equation}
where $\tilde{X}_1, \tilde{Z}_1$ correspond to measurements of the logical information for the output states.

\subsubsection{Duplex quantum transduction ($\eta=1/5$)}
For $\eta=1/5$, we have $n=5$ and $(d_1,d_2) = (2,2)$, which leads to $(d_3,d_4)=(10,10)$.
In this scenario, one qubit of quantum information can be transmitted simultaneously through $\mathcal{E}_1$ and $\mathcal{E}_2$, realizing perfect duplex quantum transduction~\cite{wang2023}. The output states (Eq.~(\ref{eq:output_states})) are (Fig.~\ref{fig4}(d))
\begin{equation}
    \begin{split}
        \ket{\psi_{0,0}} =& \sum_{j=0}^{4} \ket{4j} \ket{2j} \\
        \ket{\psi_{1,0}} =& \sum_{j=0}^{4} \ket{4j+5} \ket{2j} \\
        \ket{\psi_{0,1}} =& \sum_{j=0}^{4} \ket{4j} \ket{2j+5} \\
        \ket{\psi_{1,1}} =& \sum_{j=0}^{4} \ket{4j+5} \ket{2j+5} ,
    \end{split}
\end{equation}
where we have omitted the module 10 on the right hand side.

We can choose $\alpha_1=\alpha_2=\beta_1=\beta_2 = 1$ and the single mode logical operators (Eq.~(\ref{eq:output_logical_ops_single})) are $\tilde{X}_1 = (\hat{X}_1)^{\sqrt{5}}, \tilde{Z}_1 = (\hat{Z}_1)^{\sqrt{5}}, \tilde{X}_2 = (\hat{X}_2)^{\sqrt{5}}, \tilde{Z}_2 = (\hat{Z}_2)^{\sqrt{5}}$.
For a square lattice $S$, we have
\begin{equation}
    \begin{split}
        \tilde{X}_1 =& \exp \left( -i\sqrt{5\pi} \opp_1 \right) , \quad \tilde{Z}_1 = \exp \left( i\sqrt{5\pi} \opq_1 \right) , \\
        \tilde{X}_2 =& \exp \left( -i\sqrt{5\pi} \opp_2 \right) , \quad \tilde{Z}_2 = \exp \left( i\sqrt{5\pi} \opq_2 \right) .
    \end{split}
\end{equation}

While our primary focus is on sending quantum information, it is worth noting that our results also apply to several applications in classical communication. For instance, the entanglement generation at $\eta=1/2$ supports quantum key distribution~\cite{bennett2014,ekert1991}. The unidirectional quantum transduction at $\eta=1/3$ enables quantum secure direct communication~\cite{deng2003,sheng2022}, allowing classical messages to be sent without shared secret keys. Moreover, the duplex quantum transduction at $\eta=1/5$ leads to quantum dialogue~\cite{nguyen2004,zhu2024}, enabling simultaneous exchange of classical information without secret keys.

\section{Quantum communication with finite-energy GKP codes}
\label{sec4}
Although perfect transmission is possible using ideal GKP codes with infinite energy, in practice it is important to consider finite energy constraints.
At finite energy, we do not need to set $\eta$ precisely at some value which allows for a finite precision on the control of $\eta$.
Here we show that high fidelity quantum communication through $\mathcal{E}_1$ is achievable with finite-energy GKP codes.

\subsection{Performance of a single input encoding}
\begin{figure}[t]
	\centering
	\includegraphics[width=0.48\textwidth]{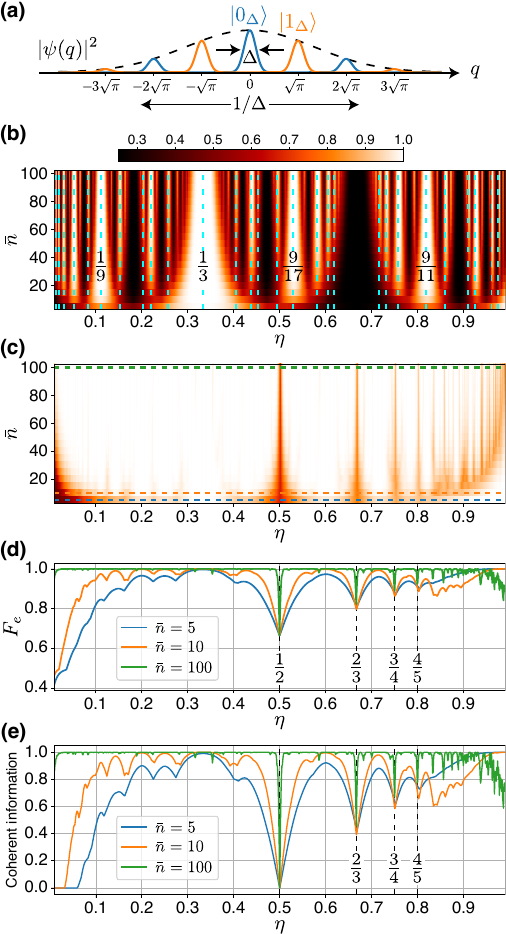}
	\caption{(a) The square lattice GKP states (Eq.~(\ref{eq:square_lattice_GKP_d2})) at finite energy. (b) Entanglement fidelity $F_e (\eta,\bar{n},\mathcal{C})$ for the input encoding $\mathcal{C} = \mathcal{C}_{d_1,S_1} \otimes \mathcal{C}_{d_2,S_2}$ with dimensions $(d_1,d_2)=(2,1)$ and square lattices $S_1=S_2=I_2$. Cyan dashed lines are the predicted peak locations from Eq.~(\ref{eq:peak_predictions}). (c) The maximal entanglement fidelity $F_e^*(\bar{n},\eta)$ over many input encodings. (d) Slices of (c) at $\bar{n}=5,10,100$. (e) The corresponding coherent information for slices of (c) at $\bar{n}=5,10,100$. For (b-e), $\eta$ ranges from 0.01 to 0.99.}
	\label{fig5}
\end{figure}

By applying the envelope operator $\exp \left( -\Delta^2 \opn \right)$ to an ideal GKP state $\ket{\mu}$, we can define a finite-energy GKP state~\cite{gottesman2001,noh2019} as
\begin{equation}
    \label{eq:finite_energy_GKP}
    \ket{\mu_{\Delta}} = \mathcal{N}_{\Delta,\mu} \exp \left( -\Delta^2 \opn \right) \ket{\mu} ,
\end{equation}
where $\mathcal{N}_{\Delta,\mu}$ is the normalization constant. The average photon number $\bar{n}$ of $\ket{\mu_{\Delta}}$ is approximately $\bar{n} \approx \frac{1}{2\Delta^2} - \frac{1}{2}$. In Fig.~\ref{fig5}(a) we plot the position space wavefunctions for the finite-energy square lattice GKP code (Eq.~(\ref{eq:square_lattice_GKP_d2})) with $\bar{n}\approx 5$. Here the width of the overall Gaussian envelope scales as $1/\Delta$ and the width of each peak scales as $\Delta$.

GKP states with a Gaussian envelope have been demonstrated experimentally for microwave photons~\cite{sivak2023} and phonons~\cite{deneeve2022}.
In optics, however, a Gaussian envelope may not be the most convenient choice although arbitrary envelope functions can be synthesized~\cite{vasconcelos2010,weigand2018,eaton2019,su2019,hastrup2022,takase2023}.
We consider binomial envelopes originating from cat states breeding~\cite{vasconcelos2010,weigand2018} in Appendix~\ref{sec:SI_other_envelopes}, and find no significant impact on the fidelity.
Intuitively, the erasure of the logical information from the environment (Fig.~\ref{fig2}(c)) still works as long as the GKP states have enough peaks with smooth envelopes.

One useful property to notice is that when the two input modes have the same $\bar{n}$, i.e., the same $\Delta$, the envelope operator $\exp \left( -\Delta^2 (\opn_1 +\opn_2) \right)$ commutes with $\opu_\eta$ since $[\opn_1+\opn_2, \opu_\eta]=0$. Therefore the peak widths of input and output states are both $\Delta$.
For an input encoding that satisfies the lattice matching condition (Eq.~(\ref{eq:lattice_matching})), the entanglement fidelity approaches 1 as $\bar{n} \rightarrow \infty$.
On the other hand, the entanglement fidelity drops at finite $\bar{n}$, since the peaks can overlap at the output which prevents a perfect decoding of the logical information.

We examine the performance of an input encoding $\mathcal{C}$ by subjecting it to beam splitters with different transmissivities $\eta$ and varying the average photon number $\bar{n}$ of the encoding.
We choose the input encoding $\mathcal{C}$ as $\mathcal{C}_{d_1,S_1} \otimes \mathcal{C}_{d_2,S_2}$ with dimensions $(d_1,d_2)=(2,1)$ and square lattices $S_1=S_2=I_2$, and its entanglement fidelity $F_e(\eta,\bar{n},\mathcal{C})$ (Fig.~\ref{fig5}(b)) shows many peaks besides $\eta=1/3$.
This is expected since the input encoding actually satisfies the lattice matching condition (Eq.~(\ref{eq:lattice_matching})) for any transmissivities
\begin{equation}
    \label{eq:peak_predictions}
    \eta = \frac{(2a+1)^2}{(2a+1)^2 + 2 b^2} ,
\end{equation}
where integers $a,b$ satisfy $\gcd(b,2a+1)=1$ and $a \geq 0, b \geq 1$, leading to the appearance of multiple peaks at $\eta=\frac{1}{3}, \frac{9}{11}, \frac{1}{9}, \frac{9}{17},...$ as $\bar{n}$ increases.
The predictions from Eq.~(\ref{eq:peak_predictions}) are shown as cyan dashed lines in Fig.~\ref{fig5}(b), which agree well with the actual peak locations.

Another feature of the $F_e(\eta,\bar{n},\mathcal{C})$ plot (Fig.~\ref{fig5}(b)) is that the width of any specific peak gets narrower at large $\bar{n}$ (for example the peak corresponding to $\eta=1/3$).
Intuitively, this is because at large $\bar{n}$, the position space wavefunctions consist of narrower peaks further away from the origin and the entanglement fidelity is thus more sensitive to small changes of $\eta$.

Furthermore, the bright and dark regions are roughly symmetric about $\eta=0.5$, e.g., $\eta=1/3$ is bright with high fidelity while $\eta=2/3$ is dark with low fidelity.
This can be explained from the competition between a quantum channel and its complementary channel.
When the quantum channel $\mathcal{E}_1$ from $\opa_1$ to $\opa_3$ has a transmissivity $\eta$, its complementary channel $\tilde{\mathcal{E}}_1$ from $\opa_1$ to $\opa_4$ is also a quantum channel with transmissivity $1-\eta$. Therefore, when $\mathcal{E}_1$ with transmissivity $\eta$ transmits quantum information with high fidelity, its complementary channel $\tilde{\mathcal{E}}_1$ with transmissivity $1-\eta$ must have a low fidelity due to no-cloning theorem.

\subsection{Performance of many input encodings}
As shown in Fig.~\ref{fig5}(b), a single input encoding can generate many high fidelity regions with low fidelity gaps in between. A natural question arises: can we achieve high fidelity at every transmissivity $\eta$ using different GKP codes?
This may seem obvious since for any rational $\eta$ (except for $\eta=(n-1)/n$), there will be a high fidelity region around it achieved with the GKP code from Eq.~(\ref{eq:lattice_matching}) at large enough $\bar{n}$, and rational numbers are dense.
However, although at large $\bar{n}$ more rational $\eta$ are covered by high fidelity regions, each region gets narrower and it is not obvious that these regions will form a dense cover as $\bar{n} \rightarrow \infty$.
After all, irrational $\eta$ require different encoding constructions (Eq.~(\ref{eq:irrational_GKP})) instead of the usual GKP codes to achieve perfect transmission.

We address this question by numerically calculating the entanglement fidelity for many input encodings. The maximal entanglement fidelity $F_e^*(\bar{n},\eta) \equiv \max_{\mathcal{C}} F_e(\bar{n},\eta,\mathcal{C})$ is shown in Fig.~\ref{fig5}(c), with three slices at $\bar{n}=5,10,100$ shown in Fig.~\ref{fig5}(d).
As we increase $\bar{n}$, high fidelity transmission is possible at nearly any $\eta$, except for $\eta=(n-1)/n$.
At $\eta \approx 1$, increasing $\bar{n}$ leads to lower fidelity since the GKP codes from Eq.~(\ref{eq:lattice_matching}) for $\eta \approx 1$ have large output dimensions and thus require even larger $\bar{n}$ to achieve high fidelity.

The maximal entanglement fidelity at a given $\bar{n}$ does not monotonically increase with $\eta$, but has a rather complicated behavior. For example, $\eta=1/3$ appears easier for achieving quantum communication than other transmissivities.

To make a fair comparison between GKP codes with different dimensions, we only consider sending a qubit with basis states $\{\ket{0_\Delta}, \ket{\lfloor d_1/2 \rfloor_\Delta }\}$ through $\mathcal{E}_1$ instead of the full $d_1$-dimensional GKP code $\mathcal{C}_{d_1,S_1}$. The environment state of $\opa_2$ is chosen as the qunaught state $\ket{0_\Delta}$ from $\mathcal{C}_{d_2,S_2}$ with $d_2=1$.
We use the transpose channel decoder to obtain good estimations of the entanglement fidelity and apply an efficient method to simulate the beam splitter with finite-energy GKP input states up to $\bar{n}=100$ in both modes. We also consider the general case where the two modes have different average photon numbers. See Appendix~\ref{sec:SI_simulation_details} for more details.

\subsubsection{Theoretical explanation}
Numerical results show that high fidelity transmission is possible for general $\eta$ at large $\bar{n}$. Here we provide a theoretical argument of why finite-energy GKP codes work well for irrational $\eta$.

With finite-energy GKP states, the fidelity will not change much for a small difference in $\eta$.
As a result, the GKP code that achieves high fidelity for a transmissivity $m/n$ may also apply to $\eta \approx m/n$ with high fidelity.
For any given $\eta$, we can transmit a $d$-dimensional subspace with high fidelity if a good rational approximation $m/n \approx \eta$ exists satisfying $d \mid (n-m)$ and
\begin{equation}
    \label{eq:irrational_requriement}
	\left| \eta - \frac{m}{n} \right| < \frac{\varepsilon}{n (\log_2 n + c)} ,
\end{equation}
where $\varepsilon \ll 1$ and $c>0$ are some constants.
This result is derived in Appendix~\ref{sec:SI_finite_energy}, assuming that the higher order terms in Eq.~(\ref{eq:appendix_assumption}) are negligible.

It has been proven in number theory~\cite{uchiyama1980} that for irrational $\eta$ and fixed integer $d$, there exist infinitely many pairs of integers $m,n$ satisfying
\begin{equation}
	\label{eq:uchiyama_theorem}
	\left| \eta - \frac{m}{n} \right| < \frac{d^2}{4n^2}
\end{equation}
where $m \equiv n \equiv 1 \pmod{d}$.
For any given $\varepsilon,c$, we have
\begin{equation}
	\frac{d^2}{4n^2} < \frac{\varepsilon}{n (\log_2 n + c)}
\end{equation}
when $n$ is large enough. Therefore, we can always find $m/n$ satisfying Eq.~(\ref{eq:irrational_requriement}) and $d \mid (n-m)$, which achieve high fidelity transmission for irrational $\eta$.

On the other hand, rational approximations of rational $\eta$ can only be bounded by $|\eta - m/n| < O(1/n)$ instead of $O(1/n^2)$ and our arguments above for irrational $\eta$ does not apply here.
This may explain the lower fidelities at $\eta=(n-1)/n$ in numerical simulations, since those transmissivities only support $d=1$ and we cannot ensure good rational approximations for them.

\subsection{Coherent information and entanglement fidelity with finite-energy GKP codes}
In addition to entanglement fidelity, we calculate the coherent information which directly establishes a lower bound on the quantum capacity. In Fig.~\ref{fig5}(e), we plot the coherent information corresponding to the slices of Fig.~\ref{fig5}(c), demonstrating that large coherent information is achievable for $\eta < 0.5$.

Furthermore, the infidelity at large $\bar{n}$ entirely comes from the non-orthogonality of finite-energy GKP states, which scales approximately as $e^{-\bar{n}}/\bar{n}$ (see Appendix~\ref{appendix:fidelity_finite_energy_GKP}). Consequently, arbitrarily small infidelity $\epsilon=1-F_e$ can be attained by choosing $\bar{n} \sim -\log \epsilon$.

\section{Discussion and Outlook}
\label{sec5}
We study quantum communication through a beam splitter channel and find that GKP states appear as the optimal input states from numerical optimization. We develop a general theory for constructing the system encoding and environment state, which shows that simultaneous bidirectional quantum communication through a beam splitter is achievable with GKP states.
We also consider finite energy effects and demonstrate that high fidelity quantum communication is possible at nearly any $\eta$ as we increase the average number of photons in the GKP states.

We propose quantum transduction as a natural application of passive environment assistance.
Quantum transducers facilitate coherent conversion of quantum states between distinct physical systems and are essential for exchanging quantum information in hybrid quantum networks~\cite{kimble2008}.
A notable example is microwave-optical quantum transducers~\cite{lauk2020,lambert2020a,han2021}, which enable the interconnection of remote superconducting qubits via optical fibers.
Most quantum transducers engineer an effective beam splitter interaction between two bosonic modes~\cite{han2021}, as illustrated in Fig.~\ref{fig1}(d). The conversion efficiency $\eta$ of the transducer corresponds to the transmissivity of the beam splitter.
We can neglect internal losses by assuming large external coupling rates for example, and model a quantum transducer as a beam splitter unitary $\opu_\eta$ (Fig.~\ref{fig1}(b)).
The transduction channels $\mathcal{E}_1$ and $\mathcal{E}_2$ convert the input modes $\opa_1$ and $\opa_2$ of mode 1 and mode 2 to the output modes $\opa_3$ and $\opa_4$ of mode 2 and mode 1 (Fig.~\ref{fig1}(d))~\cite{wang2023}.
The environment associated with the transduction channel $\mathcal{E}_1 (\mathcal{E}_2)$ is simply the idle input mode $\opa_2 (\opa_1)$.

Low conversion efficiency is one major challenge in quantum transduction due to limited coupling strength and internal losses.
The usual treatment of $\mathcal{E}_1$ and $\mathcal{E}_2$ as bosonic pure-loss channels leads to the stringent requirement of $\eta > 0.5$ for direct quantum transduction.
The efficiency requirement can be relaxed with alternative transduction schemes based on adaptive control~\cite{zhang2018}, shared entanglement~\cite{zhong2020a,krastanov2021,wu2021} or interference~\cite{lau2019,zhang2022a}. However, these methods require additional resources such as measurement and feedback which are also challenging experimentally.
The passive environment-assisted quantum transduction is the first method that can achieve direct quantum transduction even when the conversion efficiency is below 0.5.
Although the adaptive control method~\cite{zhang2018} as an example of active environment assistance has been studied, passive environment assistance remains unexplored for quantum transducers.
Environment assistance, both active and passive, suggests different transducer optimizations. Instead of maximizing the conversion efficiency, it may be advantageous to minimize the internal losses to the uncontrollable environments and then overcome the potentially lower conversion efficiencies with environment assistance.

In Appendix~\ref{appendix:intrinsic_loss}, we numerically analyze the effects of intrinsic loss  and demonstrate the benefits of our scheme even in presence of intrinsic loss.
With a few percent intrinsic loss, high fidelity and large coherent information is still achievable using GKP encoding and environment states. For $\eta=1/3$, our tri-convex optimization shows that the coherent information, and thus the quantum capacity, remains positive for 27\% intrinsic loss probability and the optimized results are still GKP states.
Experimentally, intrinsic loss below 1\% have been demonstrated in microwave-microwave quantum transduction~\cite{abdo2013}, and intrinsic loss about 20\% have been demonstrated in optical-optical quantum transduction~\cite{pelc2012}.

Currently, single-mode GKP states of microwave photons and phonons have been successfully generated on superconducting circuits~\cite{campagne-ibarcq2020,sivak2023} and ion trap platforms~\cite{fluhmann2019,deneeve2022}, with active exploration in the optical domain~\cite{brady2024,konno2024,vasconcelos2010,weigand2018,eaton2019,su2019,hastrup2022,takase2023,winnel2024}.
For microwave-optical quantum transduction, alternative encodings such as the one described in Eq.~(\ref{eq:other_encodings}) may offer advantages. With this approach, GKP states are only required on the microwave side, while the optical side can utilize squeezed cat states, which are simpler to generate compared to GKP states~\cite{winnel2024}.

In future works, it would be interesting to experimentally demonstrate quantum communication through a beam splitter with GKP states, and theoretically understand the environment-assisted quantum capacity of a quantum channel with energy constraints.
For quantum transduction, it is important to consider internal losses and assess the performance of environment-assisted quantum transduction in more realistic settings.
Moreover, we can explore alternative encodings tailored for current experimental platforms or adapt transducer designs to align with specific encodings.

\begin{acknowledgments}
	We thank Cheng Guo, Bikun Li, Guo Zheng for helpful discussions. We acknowledge support from the ARO(W911NF-23-1-0077), ARO MURI (W911NF-21-1-0325), AFOSR MURI (FA9550-19-1-0399, FA9550-21-1-0209, FA9550-23-1-0338), DARPA (HR0011-24-9-0359, HR0011-24-9-0361), NSF (OMA-1936118, ERC-1941583, OMA-2137642, OSI-2326767, CCF-2312755), NTT Research, Packard Foundation (2020-71479), and the Marshall and Arlene Bennett Family Research Program. This material is based upon work supported by the U.S. Department of Energy, Office of Science, National Quantum Information Science Research Centers.
\end{acknowledgments}

\appendix

\section{Tri-convex optimization results}
\label{sec:SI_data}
\begin{figure*}[t]
	\centering
	\includegraphics[width=0.98\textwidth]{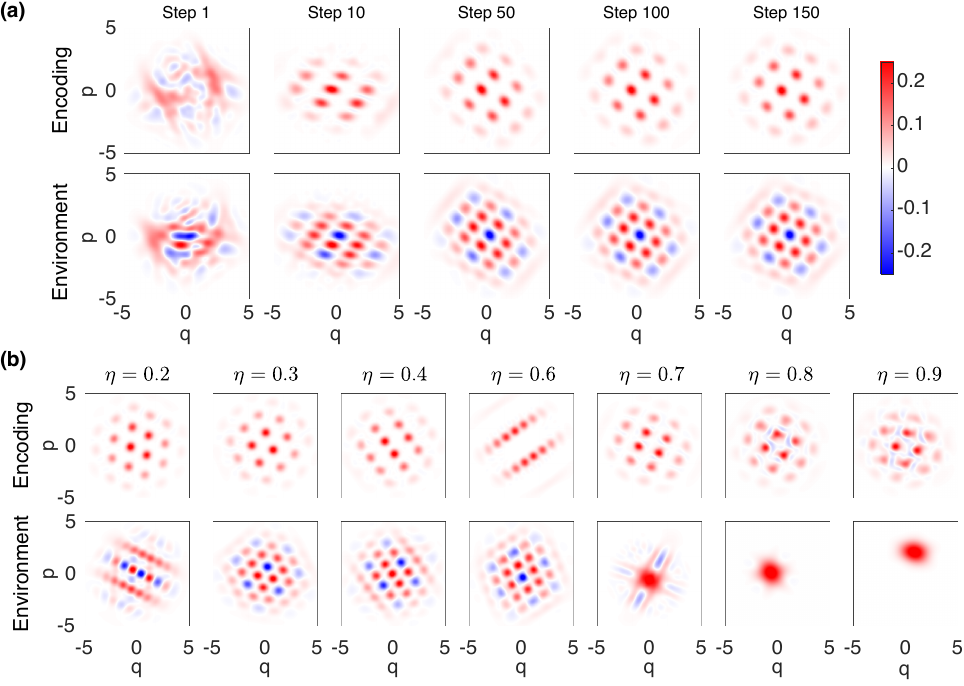}
	\caption{(a) The intermediate results during the optimization process for $\eta=1/3$, which converges after 150 steps to the results shown in Fig.~\ref{fig2}(b). (b) Tri-convex optimization results for a few other $\eta$ with entanglement fidelities 0.91, 0.97, 0.93, 0.92, 0.93, 0.98, 0.998 respectively.}
	\label{SI_fig_eta_sweep}
\end{figure*}

We plot the tri-convex optimization results for a few other $\eta$ in Fig.~\ref{SI_fig_eta_sweep}(b) subject to the average photon number constraint $\bar{n} \leq 3$. Intuitively, at large $\eta$ the input GKP encodings that satisfy the lattice matching condition Eq.~(\ref{eq:lattice_matching}) have small lattice spacings and therefore require more photons to achieve high fidelity. This may explain why at large $\eta$ and low $\bar{n}$ the environment state approaches a vacuum state which reduces the added noise to $\mathcal{E}_1$. See Appendix~\ref{sec:SI_n1n2} for more discussions.
We used CVX~\cite{cvx,gb08} package to solve the convex optimization problems.

\section{Proof of Eq.~(\ref{eq:QEC_conditions})}
\label{sec:SI_proof_1}
We can justify the perfect transmission condition from the quantum error correction conditions. The Kraus operators of the system are $\hat{E}_n = \mele{n}{\opu}{\psi_E}$ after tracing out the environment, where $\{\ket{n}\}$ is a set of complete basis of environment Hilbert space $\mathcal{H}_E$. To perfectly recover the encoded quantum information from the output of the system, quantum error correction requires~\cite{nielsen2012}
\begin{equation}
	\begin{split}
		C_{mn} \delta_{ij} =& \mele{\psi_j}{\hat{E}_m^\dagger \hat{E}_n}{\psi_i} = \tr \left[ \hat{E}_n \ket{\psi_i} \bra{\psi_j} \hat{E}_m^\dagger \right] \\
		=& \tr \left[ \bra{n} \opu (\ket{\psi_i} \bra{\psi_j} \otimes \ket{\psi_E} \bra{\psi_E}) \opud \ket{m} \right] \\
		=& \bra{n} \tr_S \left[ \opu (\ket{\psi_i} \bra{\psi_j} \otimes \ket{\psi_E} \bra{\psi_E}) \opud \right] \ket{m} ,
	\end{split}
\end{equation}
where $C$ is a Hermitian matrix.
We define the output environment state $\oprho_E = \sum_{mn} C_{mn} \ket{n} \bra{m}$, which gives Eq.~(\ref{eq:QEC_conditions}).

\section{Derivation of Eq.~(\ref{eq:construction})}
\label{sec:SI_proof_2}
Given that the equality in Fig.~\ref{fig3}(a) holds for $\opu_X=\hat{T}(\bm{\tilde{u}})$ and $\opu_Z=\hat{T}(\bm{\tilde{v}})$, we would like to find the GKP code that satisfies the perfect transmission condition Eq.~(\ref{eq:QEC_conditions}).
Instead of following Eq.~(\ref{eq:general_code}) which generates the basis states one by one, here we take a slightly different approach to directly obtain the stabilizers of the GKP code.

Any displacement operator from the 2D lattice $\Lambda(\bm{\tilde{u}}, \bm{\tilde{v}}) \equiv \{ \hat{T}(s\bm{\tilde{u}} + t\bm{\tilde{v}}), s,t \in \mathbb{Z} \}$ satisfies the same equality in Fig.~\ref{fig3}(a).
If we can find a $d$-dimensional GKP code $\mathcal{C}_{(\bm{u},\bm{v})}$ whose logical operators $\hat{X}$ and $\hat{Z}$ (up to some stabilizer displacements) exist in $\Lambda(\bm{\tilde{u}}, \bm{\tilde{v}})$, then $\mathcal{C}_{(\bm{u},\bm{v})}$ is perfectly preserved under $\opu$.
To see this, we can construct the subspace in Eq.~(\ref{eq:general_code}) by choosing $\ket{\psi_0}$ as the eigenstate $\ket{\mu=0}$ of $\hat{Z}$ with eigenvalue 1 and generating all basis states by applying $\hat{X}$. From Eq.~(\ref{eq:qudit_logical}), the resulting subspace is exactly $\mathcal{C}_{(\bm{u},\bm{v})}$.

The requirement that logical operators $\hat{X}$ and $\hat{Z}$ exist in $\Lambda(\bm{\tilde{u}}, \bm{\tilde{v}})$ is equivalent to solving the following equations for given $\bm{\tilde{u}}, \bm{\tilde{v}}$:
\begin{equation}
	\label{eq:lattice_eqs}
	\begin{split}
		s_1 \bm{\tilde{u}} + t_1 \bm{\tilde{v}} =& \left( \frac{1}{d} + \tilde{s}_1 \right) \bm{u} + \tilde{t}_1 \bm{v} \\
		s_2 \bm{\tilde{u}} + t_2 \bm{\tilde{v}} =& \tilde{s}_2 \bm{u} + \left( \frac{1}{d} + \tilde{t}_2 \right) \bm{v} ,
	\end{split}
\end{equation}
where $s,t,\tilde{s},\tilde{t}$ are all integers. This gives
\begin{equation}
	\begin{split}
		& (s_1 t_2 - s_2 t_1) \omega (\bm{\tilde{u}}, \bm{\tilde{v}}) \\
		=& \left[ \left( \frac{1}{d} + \tilde{s}_1 \right) \left( \frac{1}{d} + \tilde{t}_2 \right) - \tilde{s}_2 \tilde{t}_1 \right] \omega (\bm{u}, \bm{v}) \\
		=& \frac{2\pi}{d} \left[ (d \tilde{s}_1 + 1) (d \tilde{t}_2 + 1) - d^2 \tilde{s}_2 \tilde{t}_1 \right] .
	\end{split}
\end{equation}
Since $\gcd(d,(d \tilde{s}_1 + 1) (d \tilde{t}_2 + 1) - d^2 \tilde{s}_2 \tilde{t}_1)=1$, we have
\begin{equation}
    \omega (\bm{\tilde{u}}, \bm{\tilde{v}}) = 2\pi \frac{m}{kd}
\end{equation}
with $\gcd(m,kd)=1$.

For given $m,k,d$, the solutions of Eq.~(\ref{eq:lattice_eqs}) are not unique and any solution is a valid construction of the GKP code. For simplicity, we only focus on solutions with $t_1=\tilde{t}_1=s_2=\tilde{s}_2=0$, where Eq.~(\ref{eq:lattice_eqs}) reduces to
\begin{equation}
    \begin{split}
        s_1 \bm{\tilde{u}} =& \left( \frac{1}{d} + \tilde{s}_1 \right) \bm{u} \\
        t_2 \bm{\tilde{v}} =& \left( \frac{1}{d} + \tilde{t}_2 \right) \bm{v} .
    \end{split}
\end{equation}
We can verify that Eq.~(\ref{eq:construction}) is one possible solution that satisfies these two equations. Since $\gcd(m_1,d)=1$, there exist integers $x,y$ such that $m_1 x = d y + 1$ which leads to $s_1= k_1 x, \tilde{s}_1=y$, and the first equation above becomes $k_1 x \bm{\tilde{u}} = m_1 x \bm{u}$. Similarly, we can solve for $t_2,\tilde{t}_2$.

\section{Two-mode linear transformation}
\label{sec:SI_other_transformation}
In the main text, we construct the input GKP encodings for a beam splitter. Here we apply the same method to other two-mode linear transformations.
Two-mode Gaussian unitary can be classified into several classes~\cite{lau2019} including identity, SWAP, beam splitter, two-mode squeezing, swapped two-mode squeezing, QND, swapped QND.
Below we will look at all classes except for identity and SWAP which are trivial.

Two-mode Gaussian unitary leads to linear input-output relation between the quadratures
\begin{equation}
    \begin{pmatrix} \opq_3 \\ \opp_3 \\ \opq_4 \\ \opp_4 \end{pmatrix} = 
    S
    \begin{pmatrix} \opq_1 \\ \opp_1 \\ \opq_2 \\ \opp_2 \end{pmatrix}
    ,
\end{equation}
and the form of $S$ defines the class of the unitary.

\subsection{Two-mode squeezing}
Two-mode squeezing is given by
\begin{equation}
    S = 
    \begin{pmatrix}
        \sqrt{G} & & \sqrt{G-1} & \\
        & \sqrt{G} & & -\sqrt{G-1} \\
        \sqrt{G-1} & & \sqrt{G} & \\
        & -\sqrt{G-1} & & \sqrt{G} \\ 
    \end{pmatrix} ,
\end{equation}
where $G>1$. This leads to
\begin{equation}
	\begin{split}
	    & \opu \hat{T}_1(\bm{\alpha}) \hat{T}_2(\bm{\beta}) \\ =& \hat{T}_1(\sqrt{G}\bm{\alpha} + \sqrt{G-1} \bm{\beta}^*) \hat{T}_2 (\sqrt{G}\bm{\beta} + \sqrt{G-1} \bm{\alpha}^*) \opu ,
	\end{split}
\end{equation}
for any displacement vectors $\bm{\alpha}$ and $\bm{\beta}$. Here we have defined $\bm{\alpha}^* = (\alpha_1, -\alpha_2)$.

We choose a $d_1$-dimensional GKP code $\mathcal{C}_{(\bm{u}_1, \bm{v}_1)}$ for mode 1 and a $d_2$-dimensional GKP code $\mathcal{C}_{(\bm{u}_2, \bm{v}_2)}$ for mode 2. We would like to find the conditions such that simultaneous perfect transmission of both $\mathcal{C}_{(\bm{u}_1, \bm{v}_1)}$ and $\mathcal{C}_{(\bm{u}_2, \bm{v}_2)}$ can be achieved.

When viewing mode 2 as the environment, we have
\begin{equation}
    \begin{split}
        \bm{\tilde{u}}_1 =& \sqrt{\frac{G}{G-1}} \bm{u}_2^* \\
        \bm{\tilde{v}}_1 =& \sqrt{\frac{G}{G-1}} \bm{v}_2^* .
    \end{split}
\end{equation}
Therefore, we can choose the stabilizers of mode 1 as
\begin{equation}
	\begin{split}
		\bm{u}_1 =& \frac{d_1 k_1}{m_1} \sqrt{\frac{G}{G-1}} \bm{u}_2^* \\
		\bm{v}_1 =& -\frac{d_1 k_2}{m_2} \sqrt{\frac{G}{G-1}} \bm{v}_2^* ,
	\end{split}
\end{equation}
with the requirement
\begin{equation}
	2\pi \frac{m}{kd_1} = \omega (\bm{\tilde{u}}_1, \bm{\tilde{v}}_1) = -\frac{G}{G-1} \omega (\bm{u}_2^*,\bm{v}_2^*) = \frac{G}{G-1} 2\pi d_2 .
\end{equation}
This leads to $G = m/(m-k d_1 d_2)$.

With the choice above, we can verify that mode 2 also transmits perfectly when viewing mode 1 as the environment, since
\begin{equation}
	\begin{split}
		\bm{u}_2 =& \frac{d_2 k_2}{m_2} \sqrt{\frac{G}{G-1}} \bm{u}_1^* = \frac{d_2 k_2}{m_2} \bm{\tilde{u}}_2 \\
		\bm{v}_2 =& -\frac{d_2 k_1}{m_1} \sqrt{\frac{G}{G-1}} \bm{v}_1^* = \frac{d_2 k_1}{m_1} \bm{\tilde{v}}_2 .
	\end{split}
\end{equation}

\subsection{Swapped two-mode squeezing}
The swapped two-mode squeezing is given by
\begin{equation}
    S= 
    \begin{pmatrix}
        \sqrt{G-1} & & \sqrt{G} & \\
        & -\sqrt{G-1} & & \sqrt{G} \\
        \sqrt{G} & & \sqrt{G-1} & \\
        & \sqrt{G} & & -\sqrt{G-1} \\ 
    \end{pmatrix} ,
\end{equation}
where $G>1$. This leads to
\begin{equation}
	\begin{split}
	    & \opu \hat{T}_1(\bm{\alpha}) \hat{T}_2(\bm{\beta}) \\
        =& \hat{T}_1(\sqrt{G-1}\bm{\alpha}^* + \sqrt{G} \bm{\beta}) \hat{T}_2 (\sqrt{G-1}\bm{\beta}^* + \sqrt{G} \bm{\alpha}) \opu .
	\end{split}
\end{equation}

When viewing mode 2 as the environment, we have
\begin{equation}
    \begin{split}
        \bm{\tilde{u}}_1 =& \sqrt{\frac{G-1}{G}} \bm{u}_2^* \\
        \bm{\tilde{v}}_1 =& \sqrt{\frac{G-1}{G}} \bm{v}_2^* .
    \end{split}
\end{equation}
Therefore, we can choose the stabilizers of mode 1 as
\begin{equation}
	\begin{split}
		\bm{u}_1 =& \frac{d_1 k_1}{m_1} \sqrt{\frac{G-1}{G}} \bm{u}_2^* \\
		\bm{v}_1 =& -\frac{d_1 k_2}{m_2} \sqrt{\frac{G-1}{G}} \bm{v}_2^* ,
	\end{split}
\end{equation}
with the requirement
\begin{equation}
	2\pi \frac{m}{kd_1} = \omega (\bm{\tilde{u}}_1, \bm{\tilde{v}}_1) = -\frac{G-1}{G} \omega (\bm{u}_2^*,\bm{v}_2^*) = \frac{G-1}{G} 2\pi d_2 .
\end{equation}
This leads to $G = k d_1 d_2 / (k d_1 d_2 - m)$.

With the choice above, we can verify that mode 2 also transmits perfectly when viewing mode 1 as the environment, since
\begin{equation}
	\begin{split}
		\bm{u}_2 =& \frac{d_2 k_2}{m_2} \sqrt{\frac{G-1}{G}} \bm{u}_1^* = \frac{d_2 k_2}{m_2} \bm{\tilde{u}}_2 \\
		\bm{v}_2 =& -\frac{d_2 k_1}{m_1} \sqrt{\frac{G-1}{G}} \bm{v}_1^* = \frac{d_2 k_1}{m_1} \bm{\tilde{v}}_2 .
	\end{split}
\end{equation}

\subsection{QND}
The QND is given by
\begin{equation}
    S =
    \begin{pmatrix}
        1 & & \eta & \\
        & 1 & & \\
        & & 1 & \\
        & -\eta & & 1 \\ 
    \end{pmatrix} ,
\end{equation}
where $\eta \neq 0$. This leads to
\begin{equation}
    \label{eq:QND_displacements}
	\begin{split}
	    & \opu \hat{T}_1(\bm{\alpha}) \hat{T}_2(\bm{\beta}) \\
        =& e^{-i \frac{\eta}{2} (\alpha_1 \beta_1 - \alpha_2 \beta_2)} \hat{T}_1[(\alpha_1, \alpha_2 + \eta \beta_2)] \hat{T}_2 [(\beta_1 - \eta \alpha_1, \beta_2)] \opu .
	\end{split}
\end{equation}

There are less constraints for constructing the GKP codes for QND since only one quadrature suffers from environment noise.
We make a specific choice of rectangle GKP codes $\mathcal{C}_{d,S_1}$ and $\mathcal{C}_{d,S_2}$ for mode 1 and mode 2 respectively, where the dimension $d$ can be arbitrary and
\begin{equation}
    S_1 = \begin{pmatrix} \sqrt{d/\eta} &  \\  & \sqrt{\eta/d} \end{pmatrix}, \qquad S_2 = \begin{pmatrix} \sqrt{\eta/d} &  \\  & \sqrt{d/\eta} \end{pmatrix} .
\end{equation}
Here we assume $\eta>0$ and otherwise we can let $\eta \rightarrow -\eta$ when defining $S_1,S_2$.

When viewing mode 2 as the environment, we can choose $\opu_X$ and $\opu_Z$ in Fig.~\ref{fig3}(a) as the logical operators $\hat{X}_1$ and $\hat{Z}_1$ of mode 1 which commute with $\opu$, i.e., $\opu_X'=\opu_X$ and $\opu_Z'=\opu_Z$.
To verify this, letting $\bm{\beta}=0$ in Eq.~(\ref{eq:QND_displacements}), we have
\begin{equation}
    \opu \hat{T}_1(\bm{\alpha}) = \hat{T}_1(\bm{\alpha}) \hat{T}_2 [(- \eta \alpha_1, 0)] \opu .
\end{equation}
Since $\bm{u}_1=(\sqrt{2\pi/\eta},0)$ and $\hat{T}_2 [(-\sqrt{2\pi \eta},0)]$ is the stabilizer of mode 2, we see that $\hat{X}_1$ and $\hat{Z}_1$ commute with $\opu$ for our choice of encoding of mode 2.

Similarly, we can view mode 1 as the environment and the operators $\opu_X$ and $\opu_Z$ become the logical operators $\hat{X}_2$ and $\hat{Z}_2$ of mode 2, since $\hat{T}_1 [(0,\sqrt{2\pi \eta})]$ is the stabilizer of mode 1.
Therefore, the rectangle GKP codes $\mathcal{C}_{d,S_1}$ and $\mathcal{C}_{d,S_2}$ achieve simultaneous perfect transmission with arbitrary $d$ for QND.

\subsection{Swapped QND}
The swapped QND is given by
\begin{equation}
    S= 
    \begin{pmatrix}
        -\eta & & 1 & \\
        & & & 1 \\
        1 & & & \\
        & 1 & & \eta \\ 
    \end{pmatrix} ,
\end{equation}
where $\eta \neq 0$. This leads to
\begin{equation}
	\begin{split}
	    & \opu \hat{T}_1(\bm{\alpha}) \hat{T}_2(\bm{\beta}) \\
        = & e^{-i \frac{\eta}{2} (\alpha_1 \beta_1 - \alpha_2 \beta_2)} \hat{T}_1[(\beta_1, \beta_2 - \eta \alpha_1)] \hat{T}_2 [(\alpha_1 + \eta \beta_2, \alpha_2)] \opu .
	\end{split}
\end{equation}
Notice that the information of one quadrature is lost after $\opu$. For example, $\hat{T}_1(\bm{\alpha})$ is mapped to $\hat{T}_1[(\beta_1, \beta_2 - \eta \alpha_1)]$ where the information about $\alpha_2$ cannot be recovered.
As a result, there exist no input encoding that achieves perfect transmission for swapped QND.

\section{Output states from the beam splitter channel}
\label{sec:SI_output_states}
Here we derive several results in Sec.~\ref{sec3:output_states} characterizing the output states from the beam splitter channel.

\subsection{Derivation of Eq.~(\ref{eq:output_lattice})}
The embedding GKP lattice for the output states can be derived by finding operators that stabilize the output states and at the same time only act on one of the output modes.
The stabilizers for the input GKP codes $\mathcal{C}_{d_1,S_1} \otimes \mathcal{C}_{d_2,S_2}$ are $\hat{T}_1 (\bm{u}_1), \hat{T}_1 (\bm{v}_1), \hat{T}_2 (\bm{u}_2), \hat{T}_2 (\bm{v}_2)$.
Notice that
\begin{equation}
	\begin{split}
		\hat{T}_1 \left( d_1 k_1 \bm{\tilde{u}}_1 \right) =& \hat{T}_1 \left( m_1 \bm{u}_1 \right) \\
		\hat{T}_1 \left( d_1 k_2 \bm{\tilde{v}}_1 \right) =& \hat{T}_1 \left( m_2 \bm{v}_1 \right) \\
        \hat{T}_2 \left( d_2 k_2 \bm{\tilde{u}}_2 \right) =& \hat{T}_2 \left( m_2 \bm{u}_2 \right) \\
        \hat{T}_2 \left( d_2 k_1 \bm{\tilde{v}}_2 \right) =& \hat{T}_2 \left( m_1 \bm{v}_2 \right)
	\end{split}
\end{equation}
also stabilize the input states, from Fig.~\ref{fig3}(c-d) we can commute these operators through $\opu_\eta$ which gives that the output states are stabilized by $\hat{T}_1 (\bm{u}_3), \hat{T}_1 (\bm{v}_3), \hat{T}_2 (\bm{u}_4), \hat{T}_2 (\bm{v}_4)$, where
\begin{equation}
	\begin{split}
	    \bm{u}_3 =& \frac{m_1 \bm{u}_1}{\sqrt{\eta}}, \qquad \bm{v}_3 = \frac{m_2 \bm{v}_1}{\sqrt{\eta}} \\
        \bm{u}_4 =& \frac{m_2 \bm{u}_2}{\sqrt{\eta}}, \qquad \bm{v}_4 = \frac{m_1 \bm{v}_2}{\sqrt{\eta}} .
	\end{split}
\end{equation}
Since $\omega(\bm{u}_3,\bm{v}_3) = \frac{m}{\eta} \omega(\bm{u}_1,\bm{v}_1) = 2\pi d_1 n$ and $\omega(\bm{u}_4,\bm{v}_4) = \frac{m}{\eta} \omega(\bm{u}_2,\bm{v}_2) = 2\pi d_2 n$, the marginal output states satisfy $\oprho_3 \in \mathcal{C}_{(\bm{u}_3,\bm{v}_3)} = \mathcal{C}_{d_3,S_3}$ and $\oprho_4 \in \mathcal{C}_{(\bm{u}_4,\bm{v}_4)} = \mathcal{C}_{d_4,S_4}$, and the joint output state is embedded in $\mathcal{C}_{d_3,S_3} \otimes \mathcal{C}_{d_4,S_4}$.
Here $d_3=d_1 n, d_4=d_2 n$ and
\begin{equation}
	\begin{split}
		S_3 =& \begin{pmatrix} \sqrt{m_1/m_2} & 0 \\ 0 & \sqrt{m_2/m_1} \end{pmatrix} S_1 \\ 
		S_4 =& \begin{pmatrix} \sqrt{m_2/m_1} & 0 \\ 0 & \sqrt{m_1/m_2} \end{pmatrix} S_2 \\
		=& \begin{pmatrix} \sqrt{k_2/k_1} & 0 \\ 0 & \sqrt{k_1/k_2} \end{pmatrix} S_1 .
	\end{split}
\end{equation}

\subsection{Derivation of Eq.~(\ref{eq:output_states})}
We can derive the output states from mapping of the logical operators. The logical operators before the beam splitter are
\begin{equation}
    \begin{split}
        \hat{X}_1 =& \hat{T}_1 (\bm{u}_1/d_1), \qquad \hat{Z}_1 = \hat{T}_1 (\bm{v}_1/d_1) , \\
        \hat{X}_2 =& \hat{T}_2 (\bm{u}_2/d_2), \qquad \hat{Z}_2 = \hat{T}_2 (\bm{v}_2/d_2) .
    \end{split}
\end{equation}
The logical operators after the beam splitter are $\hat{X}_1', \hat{Z}_1', \hat{X}_2', \hat{Z}_2'$ which act on the output states give the logical information, where $\hat{O}' \equiv \opu_\eta \hat{O} \opud_\eta$. Notice that the output logical operators in general act on both output modes.

We can represent the output logical operators using the ``native'' logical operators of $\mathcal{C}_{d_3,S_3} \otimes \mathcal{C}_{d_4,S_4}$ given by
\begin{equation}
    \begin{split}
        \hat{X}_3 =& \hat{T}_1 (\bm{u}_3/d_3), \qquad \hat{Z}_3 = \hat{T}_1 (\bm{v}_3/d_3) , \\
        \hat{X}_4 =& \hat{T}_2 (\bm{u}_4/d_4), \qquad \hat{Z}_4 = \hat{T}_2 (\bm{v}_4/d_4) .
    \end{split}
\end{equation}
The relation between them is
\begin{equation}
    \label{eq:logical_ops_mapping}
	\begin{split}
		\hat{X}_1' = & (\hat{X}_3)^{m_2} (\hat{X}_4)^{-k_1 d_2} \\
		\hat{Z}_1' = & (\hat{Z}_3)^{m_1} (\hat{Z}_4)^{-k_2 d_2} \\
		\hat{X}_2' = & (\hat{X}_3)^{k_2 d_1} (\hat{X}_4)^{m_1} \\
		\hat{Z}_2' = & (\hat{Z}_3)^{k_1 d_1} (\hat{Z}_4)^{m_2} ,
	\end{split}
\end{equation}
which contains all information we need to represent the output states $\ket{\psi_{\mu_1,\mu_2}}$ in the basis of $\ket{\mu_3} \ket{\mu_4}$.

With Eq.~(\ref{eq:logical_ops_mapping}), the stabilizers of the output states are
\begin{equation}
	\begin{split}
		\hat{S}_{1X}' = & (\hat{X}_3)^{m_2 d_1} (\hat{X}_4)^{-k_1 d_1 d_2} \\
		\hat{S}_{1Z}' = & (\hat{Z}_3)^{m_1 d_1} (\hat{Z}_4)^{-k_2 d_1 d_2} \\
		\hat{S}_{2X}' = & (\hat{X}_3)^{k_2 d_1 d_2} (\hat{X}_4)^{m_1 d_2} \\
		\hat{S}_{2Z}' = & (\hat{Z}_3)^{k_1 d_1 d_2} (\hat{Z}_4)^{m_2 d_2} .
	\end{split}
\end{equation}
Notice that
\begin{equation}
	\begin{split}
		(\hat{Z}_1')^{m_2} (\hat{S}_{2Z}')^{k_2} = & (\hat{Z}_3)^n \\
		(\hat{Z}_2')^{m_1} (\hat{S}_{1Z}')^{-k_1} = & (\hat{Z}_4)^n ,
	\end{split}
\end{equation}
we have
\begin{equation}
	\begin{split}
		(\hat{Z}_3)^n \ket{\psi_{\mu_1,\mu_2}} =& \opu_\eta (\hat{Z}_1)^{m_2} \ket{\mu_1} \ket{\mu_2} = e^{i \frac{2\pi}{d_1} \mu_1 m_2} \ket{\psi_{\mu_1,\mu_2}} \\
		(\hat{Z}_4)^n \ket{\psi_{\mu_1,\mu_2}} =& \opu_\eta (\hat{Z}_2)^{m_1} \ket{\mu_1} \ket{\mu_2} = e^{i \frac{2\pi}{d_2} \mu_2 m_1} \ket{\psi_{\mu_1,\mu_2}} .
	\end{split}
\end{equation}
This leads to $\ket{\psi_{\mu_1,\mu_2}} \in \mathcal{C}_{\mu_1}^{(3)} \otimes \mathcal{C}_{\mu_2}^{(4)}$, where the subspaces $\mathcal{C}_{\mu_1}^{(3)}$ and $\mathcal{C}_{\mu_2}^{(4)}$ are defined in Eq.~(\ref{eq:direct_sum}).

To calculate $\ket{\psi_{\mu_1,\mu_2}}$, we can first express $\ket{0,0}$ in the basis of $\ket{\mu_3} \ket{\mu_4}$ and then apply the logical operators in Eq.~(\ref{eq:logical_ops_mapping}).
Let
\begin{equation}
	\ket{0,0} = \sum_{j_1,j_2=0}^{n-1} c_{j_1,j_2} \ket{j_1 d_1} \ket{j_2 d_2} \in \mathcal{C}_{0}^{(3)} \otimes \mathcal{C}_{0}^{(4)} ,
\end{equation}
we have $c_{j_1,j_2} \neq 0$ if and only if $j_1 m_1 = j_2 k_2 d_2$ since $\ket{0,0}$ is stabilized by $\hat{Z}_1' = (\hat{Z}_3)^{m_1} (\hat{Z}_4)^{-k_2 d_2}$.
Therefore we can write
\begin{equation}
	\ket{0,0} = \sum_{j=0}^{n-1} c_j \ket{j k_2 d_2 d_1} \ket{j m_1 d_2} .
\end{equation}
Furthermore, $\ket{0,0}$ is also stabilized by $\hat{S}_{2X}' = (\hat{X}_3)^{k_2 d_1 d_2} (\hat{X}_4)^{m_1 d_2}$ which leads to $c_{j+1}=c_j$. Since all $c_j$ are equal, we have
\begin{equation}
	\ket{0,0} = \frac{1}{\sqrt{n}} \sum_{j=0}^{n-1} \ket{j k_2 d_2 d_1} \ket{j m_1 d_2} .
\end{equation}

Using Eq.~(\ref{eq:logical_ops_mapping}), the output states are
\begin{equation}
	\begin{split}
		\ket{\psi_{\mu_1,\mu_2}} =& \opu_\eta (\hat{X}_1)^{\mu_1} (\hat{X}_2)^{\mu_2} \ket{0} \ket{0} \\
		=& (\hat{X}_3)^{\mu_1 m_2 + \mu_2 k_2 d_1} (\hat{X}_4)^{\mu_2 m_1 - \mu_1 k_1 d_2} \ket{0,0} \\
		=& \frac{1}{\sqrt{n}} \sum_{j=0}^{n-1} \ket{\mu_1 m_2 + \mu_2 k_2 d_1 + j k_2 d_2 d_1} \otimes \\
		& \qquad \qquad \ket{\mu_2 m_1 - \mu_1 k_1 d_2 + j m_1 d_2} .
	\end{split}
\end{equation}
Naively, it appears that the marginal state $\oprho_3 (\oprho_4)$ carries information about $\mu_2 (\mu_1)$, but this is not true since $\ket{\psi_{\mu_1,\mu_2}}$ is maximally entangled in $\mathcal{C}_{\mu_1}^{(3)} \otimes \mathcal{C}_{\mu_2}^{(4)}$.
We can show this by finding an equivalent expression of $\ket{\psi_{\mu_1,\mu_2}}$.
Notice that
\begin{equation}
	\begin{split}
		(\hat{X}_1')^{m_1} (\hat{S}_{2X}')^{k_1} = & (\hat{X}_3)^n \\
		(\hat{X}_2')^{m_2} (\hat{S}_{1X}')^{-k_2} = & (\hat{X}_4)^n ,
	\end{split}
\end{equation}
we have
\begin{equation}
	\begin{split}
		\ket{\psi_{\mu_1,\mu_2}} =& (\hat{X}_3)^{\mu_1 \alpha_1 n} (\hat{X}_4)^{\mu_2 \alpha_2 n} \ket{0,0} \\
		=& \frac{1}{\sqrt{n}} \sum_{j=0}^{n-1} \ket{\mu_1 \alpha_1 n + j k_2 d_2 d_1} \ket{\mu_2 \alpha_2 n + j m_1 d_2} ,
	\end{split}
\end{equation}
where integers $\alpha_k$ satisfy $m_k \alpha_k \equiv 1 \pmod{d_k},k=1,2$.
Since $\gcd(m,n)=1$, there exist integers $a$ and $b$ such that $am + bn = 1$ and we can choose $\alpha_1 = (a+b) m_2$ and $\alpha_2 = (a+b) m_1$.

\subsection{Single mode logical operators}
The output logical operators $\hat{X}_1', \hat{Z}_1', \hat{X}_2', \hat{Z}_2'$ in general act on both output modes. We can define equivalent single mode logical operators (Eq.~(\ref{eq:output_logical_ops_single})) as
\begin{equation}
	\begin{split}
		\tilde{X}_1 \equiv & (\hat{X}_3)^{\alpha_1 n} = (\hat{X}_1')^{\alpha_1 m_1} (\hat{S}_{2X}')^{\alpha_1 k_1} \Leftrightarrow \hat{X}_1' \\
        \tilde{Z}_1 \equiv & (\hat{Z}_3)^{\beta_1 n} = (\hat{Z}_1')^{\beta_1 m_2} (\hat{S}_{2Z}')^{\beta_1 k_2} \Leftrightarrow \hat{Z}_1' \\
		\tilde{X}_2 \equiv & (\hat{X}_4)^{\alpha_2 n} = (\hat{X}_2')^{\alpha_2 m_2} (\hat{S}_{1X}')^{-\alpha_2 k_2} \Leftrightarrow \hat{X}_2' \\
		\tilde{Z}_2 \equiv & (\hat{Z}_4)^{\beta_2 n} = (\hat{Z}_2')^{\beta_2 m_1} (\hat{S}_{1Z}')^{-\beta_2 k_1} \Leftrightarrow \hat{Z}_2' ,
	\end{split}
\end{equation}
where integers $\beta_1$ and $\beta_2$ satisfy $\beta_1 m_2 \equiv 1 \pmod{d_1}$ and $\beta_2 m_1 \equiv 1 \pmod{d_2}$, and $\Leftrightarrow $ means equivalent when acting on the output states.
We can choose $\beta_1 = (a+b) m_1$ and $\beta_2 = (a+b) m_2$.

\subsection{Subsystem decomposition}
Instead of the direct sum decomposition Eq.~(\ref{eq:direct_sum}), we can also perform a subsystem decomposition of the output GKP code $\mathcal{C}_{d_3,S_3}$ and $\mathcal{C}_{d_4,S_4}$ by defining
\begin{equation}
    \label{eq:subsystem_states}
    \begin{split}
        \ket{\mu_1}_L \otimes \ket{j_1}_G \equiv & \ket{\mu_1 \alpha_1 n + j_1 k_2 d_2 d_1 \pmod{d_3}} \\
        \ket{\mu_2}_L \otimes \ket{j_2}_G \equiv & \ket{\mu_2 \alpha_2 n + j_2 m_1 d_2 \pmod{d_4}} .
    \end{split}
\end{equation}
This leads to the decompositions
\begin{equation}
    \begin{split}
        \mathcal{C}_{d_3,S_3} =& \mathcal{C}^{L}_{d_1} \otimes \mathcal{C}^{G}_{n} \\
        \mathcal{C}_{d_4,S_4} =& \mathcal{C}^{L}_{d_2} \otimes \mathcal{C}^{G}_{n} ,
    \end{split}
\end{equation}
where $\mathcal{C}^{L}_{d_1} (\mathcal{C}^{L}_{d_2})$ is the $d_1 (d_2)$-dimensional logical subsystem and $\mathcal{C}^{G}_{n}$ is the $n$-dimensional gauge subsystem.

The output states (Eq.~(\ref{eq:output_states})) can be equivalently written as
\begin{equation}
    \ket{\psi_{\mu_1,\mu_2}} = \ket{\mu_1}_L \otimes \ket{\mu_2}_L \otimes \ket{\Psi}_G ,
\end{equation}
where
\begin{equation}
    \ket{\Psi}_G = \frac{1}{\sqrt{n}} \sum_{j=0}^{n-1} \ket{j}_G \otimes \ket{j}_G 
\end{equation}
is the state of the gauge subsystems.

We can represent the output logical operators (Eq.~(\ref{eq:logical_ops_mapping})) using the logical operators of the subsystems. Notice that
\begin{equation}
    \begin{split}
        m_2 =& m_2 (am+bn) = (a+b)m_2 n - a m_2 k d_1 d_2 \\
        = & \alpha_1 n - (a k_1 m_2) k_2 d_1 d_2 ,
    \end{split}
\end{equation}
from Eq.~(\ref{eq:subsystem_states}) we have
\begin{equation}
    (\hat{X}_3)^{m_2} = X_{d_1} \otimes X_n^{-a k_1 m_2} ,
\end{equation}
where $X_{d_1}$ and $X_{n}$ are the logical operators of subsystems $\mathcal{C}^{L}_{d_1}$ and $\mathcal{C}^{G}_{n}$.
Similarly, we can derive the subsystem decompositions
\begin{equation}
    \begin{split}
        \hat{X}_1' = & X_{d_1} \otimes I_{d_2} \otimes X_{n}^{-a k_1 m_2} \otimes X_{n}^{-a k_1 m_2} \\
        \hat{Z}_1' = & Z_{d_1} \otimes I_{d_2} \otimes Z_{n}^{m_1 k_2 d_2} \otimes Z_{n}^{-m_1 k_2 d_2} \\
        \hat{X}_2' = & I_{d_1} \otimes X_{d_2} \otimes X_{n}^{-a k d_1} \otimes X_{n}^{-a k d_1} \\
        \hat{Z}_2' = & I_{d_1} \otimes Z_{d_2} \otimes Z_{n}^{-m} \otimes Z_{n}^{m} .
    \end{split}
\end{equation}
Obviously, the output logical operators act trivially on the gauge state $\ket{\Psi}_G$. Therefore, the logical operators of mode $1 (2)$ at the input is mapped to the logical operators of the first (second) subsystem at the output by the beam splitter.

\section{Binomial envelope functions}
\label{sec:SI_other_envelopes}
GKP states with binomial envelopes can be naturally generated by breeding squeezed cat states in optics~\cite{vasconcelos2010,weigand2018}. Here we compare the fidelity of different envelopes for Eq.~(\ref{eq:square_lattice_GKP_d2}) and Eq.~(\ref{eq:square_lattice_GKP_d1}) at $\eta=1/3$ and $\bar{n}=5$, where a Gaussian envelope gives an entanglement fidelity of 0.997.

We choose the logical basis states
\begin{equation}
	\begin{split}
		\ket{\mu_1=0} \propto & \sum_{k=-1}^{1} \alpha_k \hat{D}_1(2k \sqrt{\pi} / \sqrt{2}) \ket{\zeta_1} \\
		\ket{\mu_1=1} \propto & \sum_{k=-2}^{1} \beta_k \hat{D}_1((2k+1) \sqrt{\pi} / \sqrt{2}) \ket{\zeta_1} ,
	\end{split}
\end{equation}
and the modified environment state
\begin{equation}
	\ket{\mu_2=0} = \sum_{k=-1}^{1} \alpha_k \hat{D}_2(k \sqrt{2 \pi} / \sqrt{2}) \ket{\zeta_2} .
\end{equation}
Here
\begin{equation}
    \begin{split}
        \{ \alpha_{-1}, \alpha_0, \alpha_1 \} &= \{1,2,1\} \\
        \{ \beta_{-2}, \beta_{-1}, \beta_0, \beta_1 \} &= \{ 1,3,3,1 \}
    \end{split}
\end{equation}
are the binomial coefficients and $\ket{\zeta} = \hat{S}(\zeta) \ket{0}$ is a squeezed vacuum state.
The squeezing parameters $\zeta_1,\zeta_2$ are chosen to satisfy the average photon number constraints.

The resulting entanglement fidelity is 0.943 and the reduction compared to a Gaussian envelope is mostly because $\ket{\mu_1(\mu_2)=0}$ only contain 3 peaks. If we instead choose a different binomial envelop where $\ket{\mu_1(\mu_2)=0}$ have 5 peaks with $\{ \alpha_{-2}, \alpha_{-1}, \alpha_0, \alpha_1, \alpha_2 \} = \{1,4,6,4,1\}$, the entanglement fidelity becomes 0.988, much closer to the fidelity of a Gaussian envelope.

\section{Numerical simulations}
\label{sec:SI_simulation_details}
In this section, we provide details for efficiently simulating the beam splitter channel with GKP input states. We also derive the entanglement fidelity with the transpose channel decoder.

\subsection{Beam splitter with GKP input states}
\label{sec:SI_efficient_method}
For numerical efficiency, we avoid explicit construction of the beam splitter unitary $\opu_\eta$. Instead, we represent the input states in the position or momentum space whose transformations under $\opu_\eta$ are given by
\begin{equation}
	\begin{split}
		\opu_\eta \ket{q_1} \ket{q_2} =& \ket{\sqrt{\eta} q_1 + \sqrt{1-\eta} q_2} \ket{\sqrt{\eta} q_2 - \sqrt{1-\eta} q_1} \\
		\opu_\eta \ket{p_1} \ket{p_2} =& \ket{\sqrt{\eta} p_1 + \sqrt{1-\eta} p_2} \ket{\sqrt{\eta} p_2 - \sqrt{1-\eta} p_1} ,
	\end{split}
\end{equation}
where $\ket{q}$ are the position eigenstates and $\ket{p}$ are the momentum eigenstates.
We only consider rectangle lattice GKP codes in numerical simulations since they can easily be represented in the position or momentum space.
Our method generalizes the method in Ref.~\cite{terhal2020} from single mode to the beam splitter channel with two modes.
We denote a $d$-dimensional rectangle lattice GKP code as $\mathcal{C}_{d,S_{\text{rect}}}$, where
\begin{equation}
	S_{\text{rect}} = \begin{pmatrix} \sqrt{r} &  \\  & 1/\sqrt{r} \end{pmatrix} .
\end{equation}
Here $r$ is the squeezing parameter and the stabilizers are
\begin{equation}
    \begin{split}
        \opS_X =& \exp \left( -i \sqrt{2\pi r d} \opp \right) \\
        \opS_Z =& \exp \left( i \sqrt{2\pi d/r} \opq \right) .
    \end{split}
\end{equation}

\subsubsection{Simulating the beam splitter in the position space}
The ideal rectangle lattice GKP states in the position space are
\begin{equation}
	\ket{\mu} = \sum_{k=-\infty}^{\infty} \ket{\opq=\sqrt{\frac{2\pi r}{d}}(dk+\mu)} , \mu=0,...,d-1 .
\end{equation}
The input states to the beam splitter are from the GKP code $\mathcal{C}_{d_1,S_{1,\text{rect}}} \otimes \mathcal{C}_{d_2,S_{2,\text{rect}}}$ with squeezing parameters $r_1,r_2$.
The output states from the beam splitter with finite-energy GKP input states (Eq.~(\ref{eq:finite_energy_GKP})) are
\begin{widetext}
\begin{equation}
	\label{eq:BS_output_states}
	\begin{split}
		& \opu_\eta \ket{\mu_{1,\Delta}} \ket{\mu_{2,\Delta}} \\
		= & \mathcal{N}_{\Delta,\mu_1} \mathcal{N}_{\Delta,\mu_2} e^{-\Delta^2 (\opn_1 + \opn_2)} \opu_\eta \sum_{k_1,k_2} \ket{\opq_1=\sqrt{\frac{2\pi r_1}{d_1}}(d_1 k_1 +\mu_1)} \ket{\opq_2=\sqrt{\frac{2\pi r_2}{d_2}}(d_2 k_2 +\mu_2)} \\
		=& \mathcal{N}_{\Delta,\mu_1} \mathcal{N}_{\Delta,\mu_2} e^{-\Delta^2 (\opn_1 + \opn_2)} \sum_{q_1,q_2} \ket{\sqrt{\eta} q_1 + \sqrt{1-\eta} q_2} \ket{\sqrt{\eta} q_2 - \sqrt{1-\eta} q_1} ,
	\end{split}
\end{equation}
\end{widetext}
where we have used the fact that $[\opu_\eta,\opn_1+\opn_2]=0$.
Numerically, the output states can be constructed in the Fock space with the matrix elements
\begin{equation}
	\label{eq:BS_matrix_elements}
	\begin{split}
		& (\bra{n_1} \bra{n_2}) \opu_\eta (\ket{\mu_{1,\Delta}} \ket{\mu_{2,\Delta}}) \\
		=& \mathcal{N}_{\Delta,\mu_1} \mathcal{N}_{\Delta,\mu_2} e^{-\Delta^2 (n_1+n_2)} \sum_{q_1,q_2} \\
        & \Psi_{n_1}(\sqrt{\eta} q_1 + \sqrt{1-\eta} q_2) \Psi_{n_2} (\sqrt{\eta} q_2 - \sqrt{1-\eta} q_1 ) ,
	\end{split}
\end{equation}
where $\Psi_n(q) \equiv \inp{n}{\opq=q}$ is the position wavefunction of Fock states and satisfies the recursive relation~\cite{terhal2020}
\begin{equation}
	\begin{split}
		\Psi_0(q) =& \pi^{-1/4} e^{-q^2/2} \\
		\Psi_n(q) =& q \sqrt{\frac{2}{n}} \Psi_{n-1}(q) - \sqrt{\frac{n-1}{n}} \Psi_{n-2}(q) .
	\end{split}
\end{equation}

We find it enough to choose a Fock state cutoff at $N=8/\Delta^2$. Since $\Psi_n(q)$ is negligible for $|q| > 2\sqrt{n}$, only finite number of peaks are required to construct a finite energy GKP states.
More specifically, we only need to include peaks within $|q| \leq 2\sqrt{N}$ to calculate the single mode GKP states $\ket{\mu_\Delta}$. To calculate the output states of the beam splitter, we choose a position cutoff $2\sqrt{2N}$ for both $|q_1|$ and $|q_2|$ in Eq.~(\ref{eq:BS_output_states}).

\subsubsection{Simulating the beam splitter in the momentum space}
For small $r$, the peak spacing in the position space becomes small and there are many peaks within $|q| \leq 2\sqrt{2N}$, making the summation in Eq.~(\ref{eq:BS_output_states}) less efficient numerically.
In this case, we simulate the beam splitter in the momentum space instead.

The rectangle lattice GKP states in the momentum space are
\begin{equation}
	\ket{\mu} = \sum_{k=-\infty}^{\infty} \exp \left( -i \frac{2\pi}{d} k \mu \right) \ket{\opp= k \sqrt{\frac{2\pi}{rd}}} .
\end{equation}
The output states from the beam splitter are
\begin{widetext}
\begin{equation}
	\begin{split}
		& \opu_\eta \ket{\mu_{1,\Delta}} \ket{\mu_{2,\Delta}} \\
		= & \mathcal{N}_{\Delta,\mu_1} \mathcal{N}_{\Delta,\mu_2} e^{-\Delta^2 (\opn_1 + \opn_2)} \opu_\eta \sum_{k_1,k_2} \exp \left( -i \left(\frac{2\pi}{d_1} k_1 \mu_1 + \frac{2\pi}{d_2} k_2 \mu_2 \right) \right) \ket{\opp_1=k_1 \sqrt{\frac{2\pi}{r_1 d_1}}} \ket{\opp_2=k_2 \sqrt{\frac{2\pi}{r_2 d_2}}} \\
		=& \mathcal{N}_{\Delta,\mu_1} \mathcal{N}_{\Delta,\mu_2} e^{-\Delta^2 (\opn_1 + \opn_2)} \sum_{p_1,p_2} \exp \left( -i \left(\frac{2\pi}{d_1} k_1 \mu_1 + \frac{2\pi}{d_2} k_2 \mu_2 \right) \right) \ket{\sqrt{\eta} p_1 + \sqrt{1-\eta} p_2} \ket{\sqrt{\eta} p_2 - \sqrt{1-\eta} p_1} .
	\end{split}
\end{equation}
\end{widetext}
The matrix elements are

\newpage

\begin{widetext}
\begin{equation}
	\begin{split}
		& (\bra{n_1} \bra{n_2}) \opu_\eta (\ket{\mu_{1,\Delta}} \ket{\mu_{2,\Delta}}) \\
		=& \mathcal{N}_{\Delta,\mu_1} \mathcal{N}_{\Delta,\mu_2} e^{-\Delta^2 (n_1+n_2)} \sum_{p_1,p_2} \exp \left( -i \left(\frac{2\pi}{d_1} k_1 \mu_1 + \frac{2\pi}{d_2} k_2 \mu_2 \right) \right) \tilde{\Psi}_{n_1}(\sqrt{\eta} p_1 + \sqrt{1-\eta} p_2) \tilde{\Psi}_{n_2} (\sqrt{\eta} p_2 - \sqrt{1-\eta} p_1 ) ,
	\end{split}
\end{equation}
\end{widetext}
where $\tilde{\Psi}_n(p) \equiv \inp{n}{\opp=p}$ is the momentum wavefunction of Fock states

We can relate the momentum wavefunction $\tilde{\Psi}_n(p)$ to the position wavefunction $\Psi_n(q)$.
Notice that the operator $\hat{R}= e^{i\opn \frac{\pi}{2}}$ rotates the phase space counterclockwise by $\pi/2$. As a result, it maps position eigenstate to momentum eigenstate $\hat{R} \ket{\opq=a} = \ket{\opp=a}$, or equivalently $\hat{R}^\dagger \opp \hat{R} = \opq$.
As a result, we have
\begin{equation}
	\tilde{\Psi}_n(p) = \inp{n}{\opp=p} = \mele{n}{\hat{R}}{\opq=p} = i^n \Psi_n(p).
\end{equation}

\subsection{Entanglement fidelity with the transpose channel decoder}
Although maximizing the entanglement fidelity $F_e$ over the decoder $\mathcal{D}$ is a semidefinite programming problem, the optimization can still be slow for large matrices. Instead, we choose $\mathcal{D}$ as the transpose channel decoder which gives near optimal $F_e$ that can be calculated directly without any optimization.

We can encode quantum information in a $d$-dimensional subspace $\mathcal{C}$ of mode 1 with orthogonal basis $\{ \ket{\mu}, \mu=0,...,d-1 \}$. After encoding, we have a maximally entangled state between mode 1 and a $d$-dimensional reference system $R$
\begin{equation}
	\ket{\Phi} = \frac{1}{\sqrt{d}} \sum_{\mu=0}^{d-1} \ket{\mu_R} \ket{\mu} .
\end{equation}
The input state of the beam splitter is $\ket{\Psi_0} = \ket{\Phi} \ket{\psi_E}$, where $\ket{\psi_E}$ is the initial state of mode 2.
The output state after the beam splitter is $\ket{\Psi} = (\hat{I}_R \otimes \opu) \ket{\Psi_0}$, where $\hat{I}_R$ is the identity matrix for the reference system $R$.
The subsystem states are
\begin{equation}
	\begin{split}
	    \oprho_1 =& \tr_{2R} \ket{\Psi} \bra{\Psi}, \qquad \oprho_{1R} = \tr_{2} \ket{\Psi} \bra{\Psi}, \\
        \oprho_2 =& \tr_{1R} \ket{\Psi} \bra{\Psi} , \qquad \oprho_{2R} = \tr_{1} \ket{\Psi} \bra{\Psi} .
	\end{split}
\end{equation}
The entanglement fidelity for a given decoder $\mathcal{D}$ acting on mode 1 is
\begin{equation}
    F_e = \mele{\Phi}{\mathcal{D} (\oprho_{1R})}{\Phi} .
\end{equation}

The transpose channel decoder has been proven to have near optimal performance which gives a good lower bound on the optimal entanglement fidelity~\cite{ng2010}.
The Kraus operators $\{D_i\}$ for the transpose channel decoder $\mathcal{D}$ are
\begin{equation}
	D_i = P E_i^\dagger \mathcal{E}_1 (P)^{-1/2} ,
\end{equation}
where
\begin{equation}
    P = \sum_{\mu} \ket{\mu} \bra{\mu}
\end{equation}
is the projector onto the subspace $\mathcal{C}$. The channel $\mathcal{E}_1$ is
\begin{equation}
	\mathcal{E}_1 (\oprho) = \tr_2 \left[ \opu \left( \oprho \otimes \ket{\psi_E} \bra{\psi_E} \right) \opud \right] ,
\end{equation}
and its Kraus operators are
\begin{equation}
	E_i = \mele{i}{\opu}{\psi_E} ,
\end{equation}
where $\{\ket{i}\}$ is a set of complete basis for mode 2.

The entanglement fidelity for the transpose channel decoder is $F_e = \left( \lVert \varrho_2 \rVert_F \right)^2$, where $\lVert \cdot \rVert_F$ is the Frobenius norm and $\varrho_2 = \tr_{1R} [\mathcal{N}_1 \ket{\Psi} \bra{\Psi}]$ with
\begin{equation}
	\mathcal{N}_1 \equiv \mathcal{E}_1 \left( P \right)^{-1/2} = (d \oprho_1)^{-1/2} .
\end{equation}

In our simulations, we consider the entanglement fidelity for sending a qubit with $d=2$. For a GKP code $\mathcal{C}_{d_1,S_1}$, we choose $\{\ket{0_\Delta}, \ket{\lfloor d_1/2 \rfloor_\Delta }\}$ as the qubit basis states. The environment state $\ket{\psi_E}$ is chosen as the qunaught state $\ket{\varnothing_\Delta}$ from $\mathcal{C}_{d_2,S_2}$ with $d_2=1$.
Since finite-energy GKP states are not perfectly orthogonal to each other, we need to orthogonalize the output states $\opu \ket{\mu_\Delta} \ket{\varnothing_\Delta}$ before evaluating $F_e$.

The entanglement fidelity of the complementary channel can be derived similarly.
The complementary channel $\tilde{\mathcal{E}}_1$ is
\begin{equation}
	\tilde{\mathcal{E}}_1 (\oprho) = \tr_1 \left[ \opu \left( \oprho \otimes \ket{\psi_E} \bra{\psi_E} \right) \opud \right] ,
\end{equation}
whose Kraus operators are
\begin{equation}
	E_i = \mele{i}{\opu}{\psi_E} ,
\end{equation}
where $\{\ket{i}\}$ is a set of complete basis for mode 1.
Notice that $\tilde{\mathcal{E}}_1$ corresponds to a quantum channel with transmissivity $1-\eta$ and environment state $\ket{\psi_E}$. As a result, to simulate $\eta \in [0,1]$ we only need to consider $\eta \in [0,0.5]$ and calculate the entanglement fidelity for both $\mathcal{E}_1$ and $\tilde{\mathcal{E}}_1$ at each $\eta$.

We can apply the transpose channel decoder to the complementary channel $\tilde{\mathcal{E}}_1$, and the entanglement fidelity is $F_e = \left( \lVert \varrho_1 \rVert_F \right)^2$, 
where $\varrho_1 = \tr_{2R} [\mathcal{N}_2 \ket{\Psi} \bra{\Psi}]$ and
\begin{equation}
	\mathcal{N}_2 \equiv \tilde{\mathcal{E}}_1 \left( P \right)^{-1/2} = (d \oprho_2)^{-1/2} .
\end{equation}

\subsection{Other technical details}
\noindent \textbf{GKP code list.}
We generate a list of GKP codes $\mathcal{C}_{d_1,S_{1,\text{rect}}} \otimes \mathcal{C}_{d_2,S_{2,\text{rect}}}$ for mode 1 and mode 2 to calculate the optimized entanglement fidelity in Fig.~\ref{fig5}(c). We choose square lattice for mode 1 with $r_1=1$ and dimension $d_2=1$ for mode 2. From the lattice matching condition Eq.~(\ref{eq:lattice_matching}), mode 2 is a rectangle lattice with squeezing parameter $r_2 = a/b$, where $a,b$ are coprime integers. As a result, a specific code choice can be labeled as $(d_1 \geq 2,a,b)$ and the list of codes can be roughly sorted from smaller values of $(d_1,a,b)$ to larger values. We simulate the first 1000 GKP codes on the list.

\noindent \textbf{Step size for sweeping $\eta$.}
For GKP codes with a large average photon number $\bar{n}$, the entanglement fidelity$F_e$ is more sensitive to a small change of $\eta$.
We can roughly estimate the required step size in the $\eta$ sweep to achieve a smooth curve $F_e (\eta)$.
The width of the overall Gaussian envelope is about $1/\Delta$ while the width of each peak is about $\Delta$ in the position space (Fig.~\ref{fig5}(a)). The beam splitter generates a rotation in the $(q_1,q_2)$ space with a rotation angle $\theta = \arccos \sqrt{\eta}$.
We choose a step size $\delta \eta$ such that $\frac{1}{\Delta} \delta \theta \approx \Delta$, where the changes in the output state and $F_e$ are small. Since $\delta \theta = \frac{\delta \eta}{2\sqrt{\eta (1-\eta)}}$, we have $\delta \eta \approx \sqrt{\eta (1-\eta)} / \bar{n}$.
In our simulations, $\eta$ ranges from 0.01 to 0.99 with the maximal $\bar{n}=100$, and we therefore choose the step size to be $\delta \eta \approx 0.001$.

\begin{figure}[t]
	\centering
	\includegraphics[width=0.48\textwidth]{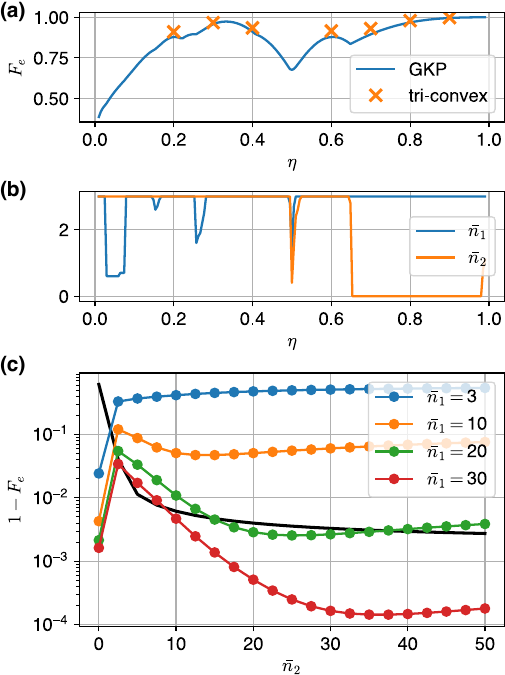}
	\caption{(a) Blue line: The performance of many input GKP encodings subject to $\bar{n}_1,\bar{n}_2 \leq 3$. Orange cross: tri-convex optimization results from Fig.~\ref{SI_fig_eta_sweep}(b). (b) The optimal values of $\bar{n}_1$ and $\bar{n}_2$ achieving the blue line in (a). (c) Infidelity as a function of $\bar{n}_2$ for $\bar{n}_1=3,10,20,30$. Here we choose $\eta=9/11$ and the input GKP encoding as $(d_1,d_2,r_1,r_2)=(2,1,1,1)$. Black line: the same input GKP encoding for $\eta=1/3$ and $\bar{n}_1=3$, where the infidelity decreases as we increase $\bar{n}_2$.}
	\label{SI_fig_n1n2}
\end{figure}

\subsection{Different mean photon numbers in two modes}
\label{sec:SI_n1n2}
In the main text, we focus on the case where the two modes have the same mean photon numbers, i.e., $\bar{n}_1=\bar{n}_2$. In principle, we can consider the more general case where $\bar{n}_1$ and $\bar{n}_2$ are different, and simulate the performance of many input encodings as in Fig.~\ref{fig5}(c).
However, since the envelope operator $\exp(-\Delta_1^2 \opn_1 - \Delta_2^2 \opn_2)$ does not commute with $\opu_\eta$, we cannot use the efficient method in Appendix~\ref{sec:SI_efficient_method} to simulate in the position or momentum space. Instead, we directly construct $\opu_\eta$ in the Fock space as a block diagonal matrix since $\opu_\eta$ preserves the total photon number in two modes.

The performance of many input GKP encodings subject to $\bar{n}_1,\bar{n}_2 \leq 3$ (blue line in Fig.~\ref{SI_fig_n1n2}(a)) is close to the tri-convex optimization results (orange cross in Fig.~\ref{SI_fig_n1n2}(a)) corresponding to Fig.~\ref{SI_fig_eta_sweep}(b).
The fidelity gap between them can be attributed to two factors: the transpose channel decoder is suboptimal and only rectangle lattices are considered.
The values of $\bar{n}_1$ and $\bar{n}_2$ (Fig.~\ref{SI_fig_n1n2}(b)) for achieving the maximal entanglement fidelities for GKP encodings gives $\bar{n}_2=0$ for large $\eta$ which is consistent with the tri-convex optimization results (Fig.~\ref{SI_fig_eta_sweep}(b)).

As we relax the average photon number constraint, the optimal environment state may transit from a vacuum state with $\bar{n}_2=0$ to a GKP state. To demonstrate this, we choose a specific input GKP encoding with $(d_1,d_2)=(2,1)$ on a square lattice and fix the beam splitter transmissivity at $\eta=9/11$ where the lattice matching condition Eq.~(\ref{eq:lattice_matching}) is satisfied.
We calculate the entanglement fidelity $F_e$ as a function of $\bar{n}_2$ for a few values of $\bar{n}_1$ in Fig.~\ref{SI_fig_n1n2}(c).
For $\bar{n}_1=3,10,20$, $\bar{n}_2=0$ is optimal, while for $\bar{n}_1=30$ the optimal environment state becomes a GKP state with $\bar{n}_2>30$.
Notice that for $\bar{n}_1=10,20,30$, local maxima of $F_e$ appear near $\bar{n}_1 \approx \bar{n}_2$. On the other hand, we simulate the same GKP code at $\eta=1/3$ and $\bar{n}_1=3$ (black line in Fig.~\ref{SI_fig_n1n2}(c)), and the entanglement fidelity $F_e$ consistently increases with $\bar{n}_2$ without any local maxima.
Understanding the optimal choices of average photon numbers will be left for future works.

\section{Derivation of Eq.~(\ref{eq:irrational_requriement})}
\label{sec:SI_finite_energy}
To derive Eq.~(\ref{eq:irrational_requriement}), we need to bound how much the performance of the quantum channel drops by some small change in $\eta$. Instead of entanglement fidelity, here we use coherent information as the performance metric which is a lower bound on the quantum capacity.
For a quantum channel $\mathcal{E}: \mathcal{L}(\mathcal{H}) \rightarrow \mathcal{L}(\mathcal{H})$, the achievable rate of quantum information with an input state $\hat{\rho}$ is measured by the coherent information $I(\mathcal{E}, \hat{\rho})$. Let $\ket{\psi} \in \mathcal{H} \otimes \mathcal{H}'$ be a purification of $\hat{\rho}$, we have
\begin{equation}
	I(\mathcal{E}, \hat{\rho}) \equiv S(\mathcal{E} (\hat{\rho})) - S ((\mathcal{E}\otimes\mathcal{I}') (\ket{\psi} \bra{\psi})), 
\end{equation}
where $S(\hat{\rho})$ is the von Neumann entropy of $\hat{\rho}$ and $\mathcal{I}'$ is the identity map on $\mathcal{H}'$.

For the quantum channel $\mathcal{E}_1$, we consider a particular input state $\oprho$ with the purification
\begin{equation}
	\ket{\Psi_0} = \frac{1}{\sqrt{d}} \left( \sum_{\mu=0}^{d-1} \ket{\mu_R} \ket{\mu_\Delta} \right) \ket{\varnothing_\Delta} ,
\end{equation}
where $\ket{\mu_\Delta}$ are the finite-energy GKP basis states for a GKP code $\mathcal{C}_{d,S}$, and the environment state of mode 2 is a GKP qunaught state.
The output state is
\begin{equation}
	\ket{\Psi_\eta} = \opu_\eta \ket{\Psi_0} ,
\end{equation}
and the coherent information is
\begin{equation}
	I(\eta) = S(\oprho_1 (\eta)) - S(\oprho_{1R} (\eta)) = S(\oprho_1 (\eta)) - S(\oprho_{2} (\eta)) ,
\end{equation}
where
\begin{equation}
	\begin{split}
		\oprho_1 (\eta) =& \tr_{2R} \ket{\Psi_\eta} \bra{\Psi_\eta}, \\
		\oprho_{1R} (\eta) =& \tr_{2} \ket{\Psi_\eta} \bra{\Psi_\eta}, \\
		\oprho_{2} (\eta) =& \tr_{1R} \ket{\Psi_\eta} \bra{\Psi_\eta} .
	\end{split} 
\end{equation}

\subsection{Bounding the coherent information}
We can bound the change of the coherent information $|I(\eta + \delta \eta) - I(\eta)|$ with the Fannes' inequality
\begin{equation}
	|S(\oprho )-S(\hat{\sigma} )|\leq 2T\log_2 (\mathbb{D})-2T\log_2 2T \equiv g(T) ,
\end{equation}
where $\mathbb{D}$ is the dimension of the density matrices and
\begin{equation}
	T (\oprho,\hat{\sigma}) = \frac{1}{2} \tr |\oprho-\hat{\sigma}|
\end{equation}
is the trace distance. Therefore
\begin{equation}
	\begin{split}
		& |I(\eta + \delta \eta) - I(\eta)| \\
		=& |S(\oprho_1 (\eta + \delta \eta)) - S(\oprho_{2} (\eta + \delta \eta)) - S(\oprho_1 (\eta)) + S(\oprho_{2} (\eta))| \\
		\leq & |S(\oprho_1 (\eta + \delta \eta)) - S(\oprho_1 (\eta))| + |S(\oprho_2 (\eta + \delta \eta)) - S(\oprho_2 (\eta))| \\
		\leq & g(T (\oprho_1 (\eta + \delta \eta), \oprho_1 (\eta))) + g(T (\oprho_2 (\eta + \delta \eta), \oprho_2 (\eta))) \\
		\leq & 2 g(T (\ket{\Psi (\eta + \delta \eta)}, \ket{\Psi (\eta)})) ,
	\end{split}
\end{equation}
where we have used the fact that trace distance is contractive under partial trace.

The beam splitter unitary is given by $\opu_\eta = \exp (-i \oph \arccos \sqrt{\eta})$, where $\oph = i(\opad_1 \opa_2 - \opa_1 \opad_2)$. 
When $\delta \eta \ll 1$, we have
\begin{equation}
	\begin{split}
		& \arccos \sqrt{\eta + \delta \eta} - \arccos \sqrt{\eta} \\
        \approx & - \frac{1}{2 \sqrt{\eta (1-\eta)}} \delta \eta + \frac{1-2\eta}{8[\eta (1-\eta)]^{3/2}} \delta \eta^2 + O(\delta \eta^3) \\
		\equiv & A_1 \delta \eta + A_2 \delta \eta^2 + O(\delta \eta^3) ,
	\end{split}
\end{equation}
which gives
\begin{equation}
    \label{eq:appendix_assumption}
	\begin{split}
		& T (\ket{\Psi (\eta + \delta \eta)}, \ket{\Psi (\eta)}) \\
        =& \sqrt{1 - |\inp{\Psi (\eta)}{\Psi (\eta + \delta \eta)}|^2} \\
		=& \sqrt{1 - \left| \mele{\Psi_0}{e^{-i \oph (A_1 \delta \eta + A_2 \delta \eta^2 + ...)}}{\Psi_0} \right|^2} \\
        =& |A_1| |\delta \eta| \sqrt{\ave{\oph^2} - \ave{\oph}^2} + O(\delta \eta^2) ,
	\end{split}
\end{equation}
where the average value is taken for $\ket{\Psi_0}$.
Bounding the higher order terms $O(\delta \eta^2)$ seems challenging since they include terms like $\ave{\oph^{n}}$ which may grow quickly with $n$.
To proceed, we make an assumption that when the first order term is small, the higher order terms are much smaller than the first order term and can be ignored.

For a rectangle lattice qunaught state $\ket{\varnothing_\Delta}$, we have $\mele{\varnothing_\Delta}{\opa_2}{\varnothing_\Delta} = 0$ and $\ave{\oph}=0$. Notice that
\begin{equation}
	\left| \ave{\opa^{\dagger 2}_1 \opa_2^2 + \opa^{\dagger 2}_2 \opa_1^2} \right| \leq \ave{\opa^{\dagger 2}_1 \opa_1^2 + \opa^{\dagger 2}_2 \opa_2^2} ,
\end{equation}
which can be derived from the fact that $\ave{\hat{A}^\dagger \hat{A}} \geq 0$ where $\hat{A}_{\pm} = \opa_1^2 \pm \opa_2^2$. Therefore
\begin{equation}
	\begin{split}
		\ave{\oph^2} =& \ave{\opn_1 \opn_2 + \opn_1 + \opn_2 - \opa^{\dagger 2}_1 \opa_2^2 - \opa^{\dagger 2}_2 \opa_1^2} \\
		\leq & \ave{\opn_1 \opn_2 + \opn_1 + \opn_2 + \opa^{\dagger 2}_1 \opa_1^2 + \opa^{\dagger 2}_2 \opa_2^2} \\
		=& \ave{\opn_1 \opn_2 + \opn_1^2 + \opn_2^2} .
	\end{split}
\end{equation}
For a finite-energy GKP state, we have the estimations $\ave{\opn} \approx \frac{1}{2\Delta^2}$ and $\ave{\opn^2} \approx \frac{1}{2\Delta^4}$ for small $\Delta$ (see below).
This gives
\begin{equation}
    \label{eq:trace_distance_bound}
	T (\ket{\Psi (\eta + \delta \eta)}, \ket{\Psi (\eta)}) \lessapprox \frac{\sqrt{5} |\delta \eta|}{4 \Delta^2 \sqrt{\eta (1-\eta)}} ,
\end{equation}
which agrees with our intuition that a larger state can only tolerate a smaller $\delta \eta$.

\subsubsection{$\ave{\opn}$ and $\ave{\opn^2}$ for finite-energy GKP states}
\label{sec:GKP_moments}
We can estimate $\ave{\opn}$ and $\ave{\opn^2}$ for finite-energy GKP states in the regime of $\Delta \rightarrow 0$. The position space wavefunction of a rectangle-lattice GKP state $\ket{\mu_\Delta}$ is
\begin{equation}
	\psi(q) = \sqrt[4]{\frac{2rd}{\pi}} \sum_{k=-\infty}^{\infty} e^{-\frac{1}{2}\Delta^2 q_k^2} e^{- \frac{(q-q_k)^2}{2\Delta^2}} ,
\end{equation}
where
\begin{equation}
	q_k = \sqrt{\frac{2\pi r}{d}}(dk+\mu) .
\end{equation}
Since $\opn = \frac{1}{2} (\opq^2+\opp^2 - 1)$ and $\opn^2 = \frac{1}{4} ((\opq^2+\opp^2)^2 - 2(\opq^2+\opp^2) + 1)$, we just need to calculate $\ave{\opq^2+\opp^2}$ and $\ave{(\opq^2+\opp^2)^2}$, which can be done in the position space.

In the regime of $\Delta \rightarrow 0$, we ignore the overlaps between different peaks for $\psi(q)$. With this approximation, we have
\begin{equation}
	\begin{split}
		& \ave{\opq^2+\opp^2} \\
        \approx & \sqrt{\frac{2rd}{\pi}} \sum_{k=-\infty}^{\infty} \\
        & e^{-\Delta^2 q_k^2} \int_{-\infty}^\infty \text{d}q \, e^{- \frac{(q-q_k)^2}{2\Delta^2}} \left( q^2 - \frac{\partial^2}{\partial q^2} \right) e^{- \frac{(q-q_k)^2}{2\Delta^2}} \\
		=& \sqrt{2rd} \sum_{k=-\infty}^{\infty} e^{-\Delta^2 q_k^2} \frac{1+ 2 q_k^2 \Delta^2 + \Delta^4}{2\Delta} \\
		\approx & \sqrt{2rd} \int_{-\infty}^{\infty} \text{d}k \, e^{-\Delta^2 q_k^2} \frac{1+ 2 q_k^2 \Delta^2 + \Delta^4}{2\Delta} \\
		=& \frac{2+\Delta^4}{2 \Delta^2} \approx \frac{1}{\Delta^2} .
	\end{split}
\end{equation}
Similarly, we can derive
\begin{equation}
    \ave{(\opq^2+\opp^2)^2} \approx \frac{2}{\Delta^4} .
\end{equation}
Therefore
\begin{equation}
	\ave{\opn} \approx \frac{1}{2\Delta^2}, \qquad \ave{\opn^2} \approx \frac{1}{2\Delta^4} 
\end{equation}
for finite-energy GKP states.

\subsection{Rational $\eta$}
When $\eta$ is rational, near perfect transmission is achievable when the width $\Delta$ of the peak is sufficiently smaller than spacing between the adjacent peaks for both input and output states.
For $\eta=m/n$ with $n-m=kd$, we choose the input GKP lattices as
\begin{equation}
	S_1 = \begin{pmatrix} \sqrt{\frac{1}{m}} & 0 \\ 0 & \sqrt{m} \end{pmatrix} , \quad
	S_2 = \begin{pmatrix} \sqrt{k} & 0 \\ 0 & \sqrt{\frac{1}{k}} \end{pmatrix}, 
\end{equation}
corresponding to $(m_1,m_2) = (m,1)$ and $(k_1,k_2) = (1,k)$ with $d_1=d,d_2=1$.
The output GKP lattices are
\begin{equation}
	S_3 = \begin{pmatrix} 1 & 0 \\ 0 & 1 \end{pmatrix}, \quad
	S_4 = \begin{pmatrix} \sqrt{\frac{k}{m}} & 0 \\ 0 & \sqrt{\frac{m}{k}} \end{pmatrix} ,
\end{equation}
with $d_3=dn,d_4=n$.

The peak spacings for $\mathcal{C}_{d_1,S_1}$, $\mathcal{C}_{d_2,S_2}$ and $\mathcal{C}_{d_3,S_3}$ are
\begin{equation}
    \sqrt{\frac{2\pi}{md}} = \sqrt{\frac{2\pi}{nd}} \frac{1}{\sqrt{\eta}} , \quad \sqrt{\frac{2\pi}{k}} = \sqrt{\frac{2\pi}{nd}} \sqrt{\frac{1}{1-\eta}} d, \quad \sqrt{\frac{2\pi}{nd}} 
\end{equation}
respectively. The peak spacing for $\mathcal{C}_{d_4,S_4}$ is
\begin{equation}
    \begin{split}
        & \min \left \{ \sqrt{\frac{2\pi k}{mn}}, \sqrt{\frac{2\pi m}{kn}} \right \} \\
        =& \sqrt{\frac{2\pi}{nd}} \min \left \{ \sqrt{\frac{1-\eta}{\eta}}, d \sqrt{\frac{\eta}{1-\eta}} \right \} .
    \end{split}
\end{equation}
Therefore, the smallest peak spacing for all input and output GKP lattices in position or momentum space can be written as $c (d,\eta)/\sqrt{n}$, where
\begin{equation}
	c (d,\eta) = \sqrt{\frac{2\pi}{d}} \min \left \{ 1, \sqrt{\frac{1-\eta}{\eta}}, d \sqrt{\frac{\eta}{1-\eta}} \right \} .
\end{equation}

To achieve near perfect transmission which leads to an achievable information rate of $I(\eta=m/n) \approx \log_2 d$, we choose
\begin{equation}
	\label{eq:Delta_choice}
	\Delta = \varepsilon_0 \frac{c (d,\eta)}{\sqrt{n}} ,
\end{equation}
where $\varepsilon_0 \ll 1$ is some small constant, such that the peaks of the input and output lattices can be well resolved.

\subsection{Rational approximations of irrational $\eta$}
When $\eta$ is irrational, we first find a rational transmissivity $m/n \approx \eta$. For the input encoding $\mathcal{C}_{d,S_1} \otimes \mathcal{C}_{1,S_2}$, we have $I(m/n) \approx \log_2 d$ with the choice of $\Delta$ from Eq.~(\ref{eq:Delta_choice}).
We can apply the same encoding to the beam splitter with irrational transmissivity $\eta$, and the difference of the coherent information can be bounded by
\begin{equation}
	\label{eq:IC_bound}
	\begin{split}
		|I(\eta) - I(m/n)| \leq & 2 (2T\log_2 (\mathbb{D})-2T\log_2 2T) \\
		<& 4 (T \log_2 (\mathbb{D}) + \sqrt{2T} ) ,
	\end{split}
\end{equation}
where we have used the Fannes' inequality and the fact that $-x \log_2 x < 2\sqrt{x}$ when $x \rightarrow 0$. Here $T$ is the trace distance $T (\ket{\Psi_\eta}, \ket{\Psi_{m/n}})$. The Fock space cutoff $\mathbb{D}$ for mode 1 can be set as $\mathbb{D} = \frac{1}{\varepsilon_1 \Delta^2}$ where $\varepsilon_1 \ll 1$ is some small constant.

Our goal is to find good approximations such that the trace distance $T$ can be sufficiently small.
More concretely, we would like to find a good approximation $m/n$ such that
\begin{equation}
    T (\ket{\Psi_\eta}, \ket{\Psi_{m/n}}) < \varepsilon_2/ \log_2 (\mathbb{D}) ,
\end{equation}
where $\varepsilon_2 \ll 1$ is a small constant. This leads to
\begin{equation}
	|I(\eta) - I(m/n)| < 4 (\varepsilon_2 + \sqrt{2 \varepsilon_2 / \log_2 (\mathbb{D})} ) \approx 0 ,
\end{equation}
which gives $I(\eta) \approx \log_2 d$ and thus high fidelity transmission of a $d$-dimensional GKP code is possible with irrational $\eta$.

From Eq.~(\ref{eq:trace_distance_bound}) and Eq.~(\ref{eq:Delta_choice}), we have
\begin{equation}
	T (\ket{\Psi_\eta}, \ket{\Psi_{m/n}}) \lessapprox n |\delta \eta| c_1 ,
\end{equation}
where $\delta \eta = \eta - \frac{m}{n}$ and
\begin{equation}
	c_1 = \frac{\sqrt{5}}{4 \varepsilon_0^2 c (d,\eta)^2 \sqrt{\eta (1-\eta)}} .
\end{equation}
Notice that $c_1$ does not depend on which approximation $m/n$ we choose since it is independent of $n$, therefore we can treat $c_1$ as a constant.

Although it is always possible to find closer approximation such that $|\delta \eta |$ is arbitrarily small, those approximations may require larger $n$ which not necessarily leads to a sufficiently small $T$. To achieve $T < \varepsilon_2/ \log_2 (\mathbb{D})$, we need to have
\begin{equation}
    \label{eq:inequality}
	|\delta \eta| < \frac{\varepsilon_2}{n ( \log_2 n + c_2 ) c_1} 
\end{equation}
where
\begin{equation}
	c_2 = - \log_2 \varepsilon_1 - 2\log_2 \varepsilon_0 - 2 \log_2 c(d,\eta) > 0 .
\end{equation}
Since $\eta,d,\varepsilon_0, \varepsilon_1, \varepsilon_2$ are all given and therefore can be treated as constants, the only variable on the right hand side of the inequality Eq.~(\ref{eq:inequality}) is $n$ which leads to Eq.~(\ref{eq:irrational_requriement}).

\begin{figure*}[t]
	\centering
	\includegraphics[width=0.98\textwidth]{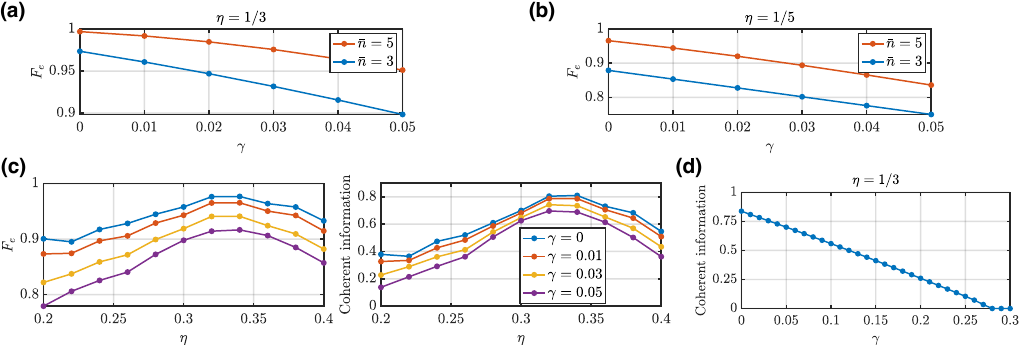}
	\caption{(a-b) Entanglement fidelity $F_e$ versus intrinsic loss probabilities $\gamma$ for transmissivities $\eta=1/3$ and $\eta=1/5$ respectively, using GKP inputs at average photon number $\bar{n}=3,5$. (c) Tri-convex optimization for $\eta \in [0.2,0.4]$ and different intrinsic loss probabilities $\gamma$ with the constraints $\bar{n} \leq 3$. We plot both the entanglement fidelity (left) and the coherent information (right) of the optimization results. (d) Tri-convex optimization at $\eta=1/3$, where the coherent information and thus the quantum capacity is positive for intrinsic loss probability $\gamma \lesssim 0.27$.}
	\label{SI_fig_intrinsic_loss}
\end{figure*}

\section{Entanglement fidelity with finite-energy GKP states}
\label{appendix:fidelity_finite_energy_GKP}
Here we would like to bound the entanglement fidelity with finite-energy GKP states, which shows that the entanglement fidelity can be arbitrarily high as we increase the average photon number $\bar{n}$ of the GKP states. We will study the specific example of $\eta=1/3$ for simplicity, but the derivation also applies to general rational $\eta$.

The finite-energy GKP states for the input subspaces $\mathcal{C}_{d_1,S_1}$ (Eq.~(\ref{eq:square_lattice_GKP_d2})) and $\mathcal{C}_{d_2,S_2}$ (Eq.~(\ref{eq:square_lattice_GKP_d1})) can be represented in the position space as
\begin{equation}
	\ket{\mu_\Delta}_1 = \sqrt[4]{\frac{4}{\pi}} \sum_{k_1=-\infty}^{\infty} e^{-\frac{1}{2}\Delta^2 q_{k_1}^2} \int \text{d}q\, e^{- \frac{(q-q_{k_1})^2}{2\Delta^2}} \ket{q} , \mu = 0,1,
\end{equation}
and
\begin{equation}
	\ket{0_\Delta}_2 = \sqrt[4]{\frac{2}{\pi}} \sum_{k_2=-\infty}^{\infty} e^{-\frac{1}{2}\Delta^2 q_{k_2}^2} \int \text{d}q\, e^{- \frac{(q-q_{k_2})^2}{2\Delta^2}} \ket{q} ,
\end{equation}
where
\begin{equation}
	q_{k_1} = (2k_1+\mu) \sqrt{\pi} , \qquad q_{k_2} = k_2 \sqrt{2\pi} .
\end{equation}

Similarly, the finite-energy GKP states for the corresponding output subspaces $\mathcal{C}_{d_3,S_3}$ and $\mathcal{C}_{d_4,S_4}$ are
\begin{equation}
	\ket{\mu_\Delta}_3 = \sqrt[4]{\frac{12}{\pi}} \sum_{k_3=-\infty}^{\infty} e^{-\frac{1}{2}\Delta^2 q_{k_3}^2} \int \text{d}q\, e^{- \frac{(q-q_{k_3})^2}{2\Delta^2}} \ket{q} , \mu = 0\sim 5,
\end{equation}
and
\begin{equation}
	\ket{\mu_\Delta}_4 = \sqrt[4]{\frac{6}{\pi}} \sum_{k_4=-\infty}^{\infty} e^{-\frac{1}{2}\Delta^2 q_{k_4}^2} \int \text{d}q\, e^{- \frac{(q-q_{k_4})^2}{2\Delta^2}} \ket{q} , \mu = 0\sim 2,
\end{equation}
where
\begin{equation}
	q_{k_3} = (6k_3+\mu) \sqrt{\frac{\pi}{3}} , \qquad q_{k_4} = (3k_4+\mu) \sqrt{\frac{2\pi}{3}} .
\end{equation}

The output states after the beam splitter are
\begin{widetext}
\begin{equation}
    \begin{split}
        & \opu_\eta \ket{\mu_\Delta}_1 \ket{0_\Delta}_2 \\
        =& \sqrt[4]{\frac{8}{\pi^2}} \sum_{k_1,k_2} e^{-\frac{1}{2}\Delta^2 (q_{k_1}^2 + q_{k_2}^2)} \iint \text{d}q_1 \text{d}q_2 \, \exp \left( -\frac{1}{2\Delta^2} [(\sqrt{\eta} q_1 - \sqrt{1-\eta} q_2 - q_{k_1})^2 + (\sqrt{\eta} q_2 + \sqrt{1-\eta} q_1 - q_{k_2})^2] \right) \ket{q_1,q_2} \\
        =& \sqrt[4]{\frac{8}{\pi^2}} \sum_{k_1,k_2} e^{-\frac{1}{2}\Delta^2 (q_{k_1}^2 + q_{k_2}^2)} \iint \text{d}q_1 \text{d}q_2 \, \exp \left( -\frac{1}{2\Delta^2} [(q_1 - (\sqrt{\eta} q_{k_1} + \sqrt{1-\eta} q_{k_2}))^2 + (q_2 - (\sqrt{\eta} q_{k_2} - \sqrt{1-\eta} q_{k_1}))^2] \right) \ket{q_1,q_2} \\
        =& \sqrt[4]{\frac{8}{\pi^2}} \sum_{k_1,k_2} e^{-\frac{1}{2}\Delta^2 (q_{k_1}^{\prime 2} + q_{k_2}^{\prime 2})} \iint \text{d}q_1 \text{d}q_2 \, \exp \left( -\frac{1}{2\Delta^2} [(q_1 - q_{k_1}')^2 + (q_2 - q_{k_2}')^2] \right) \ket{q_1,q_2} ,
    \end{split}
\end{equation}
\end{widetext}
where
\begin{equation}
    \begin{split}
        q_{k_1}' =& \sqrt{\frac{\pi}{3}} (2k_1 + 2k_2 + \mu) \\
        q_{k_2}' =& \sqrt{\frac{2\pi}{3}} (k_2 - 2k_1 - \mu) .
    \end{split}
\end{equation}
Let $k_2 - 2k_1 - \mu = 3 k_4 + \mu_4$, where $k_4 \in (-\infty, \infty)$ and $\mu_4 = 0,1,2$, we have
\begin{equation}
    2k_1 + 2k_2 + \mu = 6(k_1 + k_4) + 2 \mu_4 + 3\mu .
\end{equation}
Define $k_3 = k_1 + k_4$ and
\begin{equation}
    \begin{split}
        q_{k_3} \equiv q_{k_1}' =& \sqrt{\frac{\pi}{3}} (6k_3 + 2 \mu_4 + 3\mu) \\
        q_{k_4} \equiv q_{k_2}' =& \sqrt{\frac{2\pi}{3}} (3 k_4 + \mu_4) ,
    \end{split}
\end{equation}
we have
\begin{widetext}
\begin{equation}
    \begin{split}
        & \opu_\eta \ket{\mu_\Delta}_1 \ket{0_\Delta}_2 \\
        =& \sqrt[4]{\frac{8}{\pi^2}} \sum_{k_3,k_4 = -\infty}^\infty \sum_{\mu_4=0,1,2} e^{-\frac{1}{2}\Delta^2 (q_{k_3}^2 + q_{k_4}^2)} \iint \text{d}q_1 \text{d}q_2 \, \exp \left( -\frac{1}{2\Delta^2} [(q_1 - q_{k_3})^2 + (q_2 - q_{k_4})^2] \right) \ket{q_1,q_2} \\
        =& \frac{1}{\sqrt{3}} \left(  \ket{(3\mu)_\Delta}_3 \ket{0_\Delta}_4 + \ket{(3\mu+2)_\Delta}_3 \ket{1_\Delta}_4 + \ket{(3\mu+4)_\Delta}_3 \ket{2_\Delta}_4 \right) .
    \end{split}
\end{equation}
\end{widetext}
For small $\Delta$, the infidelity entirely comes from the non-orthogonality of finite-energy GKP states, which scales as~\cite{gottesman2001}
\begin{equation}
    1-F_e \sim \Delta^2 e^{-1/\Delta^2} \sim e^{-\bar{n}}/\bar{n} .
\end{equation}
The code-dependent constant factors are ignored here.
Therefore, arbitrarily high entanglement fidelity $F_e = 1-\epsilon$ can be achieved with $\bar{n} \sim -\log \epsilon$.

\section{Performance with intrinsic loss}
\label{appendix:intrinsic_loss}
We can add intrinsic loss to both $\opa_1$ and $\opa_2$ before and after the beam splitter, which models the lossy coupling between the input and output ports with the cavity modes. Specifically, the whole transduction process is described as $\mathcal{N}_{\sqrt{1-\gamma}}^{\otimes 2} \opu_\eta \mathcal{N}_{\sqrt{1-\gamma}}^{\otimes 2}$, where $\gamma$ is total intrinsic loss probability and $\mathcal{N}_\kappa$ represents the pure-loss channel with transmissivity $\kappa$.

We evaluate the performance of GKP input encodings in presence of intrinsic loss for both $\eta=1/3$ (Fig.~\ref{SI_fig_intrinsic_loss}(a)) and $\eta=1/5$ (Fig.~\ref{SI_fig_intrinsic_loss}(b)). The results show that the entanglement fidelity remains high for $\gamma$ about a few percent, which is achievable in microwave-microwave quantum transduction~\cite{abdo2013}.

We also perform tri-convex optimization for a range of transmissivities $\eta \in [0.2,0.4]$ below 50\% in presence of intrinsic loss.
High entanglement fidelity and large coherent information is still achievable (Fig.~\ref{SI_fig_intrinsic_loss}(c)).
We also perform tri-convex optimization for a wider range of $\gamma$ at $\eta=1/3$ with the constraints $\bar{n}\leq 3$, where the coherent information as a lower bound of the quantum capacity remains positive for $\gamma \lesssim 0.27$ (Fig.~\ref{SI_fig_intrinsic_loss}(d)). The encoding and environment states from all tri-convex optimizations are consistently GKP states, which demonstrates the robustness of our scheme against intrinsic loss. Notably, intrinsic loss probability about 20\% is achievable in optical-optical quantum transduction~\cite{pelc2012}.

% \nocite{*}
\bibliography{GKP_transduction}

\end{document}